\documentclass[aps,pre,reprint,superscriptaddress,longbibliography 
]{revtex4-1}

\usepackage{graphicx,epstopdf}
\usepackage{dcolumn}
\usepackage{bm}
\usepackage{float} 
\usepackage{enumerate}
\usepackage[sort&compress]{natbib}
\usepackage{xcolor}
\usepackage{adjustbox}

\begin{document}


\title{Molecular dynamics of partially confined Lennard-Jones gases : Velocity autocorrelation function, Mean squared displacement and collective excitations}


\author{Kanka Ghosh}
\altaffiliation{kankaghosh@physics.iitm.ac.in}
\affiliation{Department of Physics,Indian Institute of Technology Madras,Chennai-600036,India}
\author{C.V.Krishnamurthy}
\email{cvkm@iitm.ac.in}
\affiliation{Department of Physics,Indian Institute of Technology Madras,Chennai-600036,India}



\begin{abstract}
Particle motion and correlations in fluids within confined domains promise to provide challenges and opportunities for experimental and theoretical studies. We report molecular dynamics simulations of a Lennard-Jones gas mimicking argon under partial confinement for a wide range of densities at a temperature of $300$K. The isotropic behaviour of velocity autocorrelation function (VACF) and mean squared displacement (MSD), seen in the bulk, breaks down due to partial confinement. A distinct trend emerges in the VACF$_\perp$ and MSD$_\perp$ , corresponding to the confined direction, while the trends in VACF$_\parallel$ and MSD$_\parallel$, corresponding to the other two unconfined directions are seen to be unaffected by the confinement. VACF$_\perp$ displays a minimum, at short-time scales, that correlates with the separation between the reflective walls. The effect of partial confinement on MSD$_\perp$ is seen to manifest as a transition from diffusive to sub-diffusive motion with the transition time correlating with the minimum in the VACF$_\perp$. When compared to the trends shown by MSD and VACF in the bulk, the MSD$_\perp$ exhibits subdiffusive behavior, and the VACF$_\perp$ features rapid decay, suggesting that confinement suppresses the role of thermal fluctuations significantly. Repeatitive wall mediated collisions are identified to give rise to the minima in VACF$_\perp$ and in turn a characteristic frequency in its frequency spectrum. The strong linear relation between the minima in VACF$_\perp$ and wall-spacing, suggests the existence of collective motion propagating at the speed of sound. These numerical experiments, can offer interesting possibilities in the study of confined motion with observable consequences. 

\begin{description}
\item[PACS numbers]
47.11.Mn, 05.20.Jj, 51.35.+a, 51.40.+p, 47.35.Rs
\end{description}
\end{abstract}
\maketitle

\section{\label{sec:level1}Introduction}

It is quite well known that fluids behave in an unusual manner under confinement, in contrast to its bulk counterpart \cite{Mittal2006,Mittal2008}. Not only fluids, Brownian particles also show extremely different diffusive behavior under confinement with respect to the bulk environment \cite{Benesch}. Theoretical, simulation and experimental studies further confirm qualitative and quantitatively different transport and structural properties of fluids, mostly liquids, in confined spaces unlike its bulk behaviour\cite{Lobry1996,Faucheux1994,Vanderlick1987,Krishnan2003,Lancon2002,Magda1985,Pagonabarraga1999,Schoen1987,Schoen1988,Liu,Suvendu,Vadhana}. Confined gases, on the other hand, though less explored, offer a variety of unusual consequences which are distinct from both bulk gas and liquids. Effect of characteristic length and system boundaries on the mean free path of confined gases have been studied by Sree Hari et al. \cite{sree2015,Sooraj2013} using molecular dynamics(MD) simulation. At low density gases, under confinement, significant effect of characteristic length on the mean free path of the gas molecules has been found. The variation of gas molecular mean free path in nanopores with different gas-wall interaction strength has also been studied via MD \cite{Qixin2014}. MD simulations are also employed to study the anisotropic stress variation for dilute and dense confined gases \cite{Murat2011}. Markvoort et al. \cite{Markvoort2005} carried out studies of the influence of wall-gas interactions on heat flow in micro-channel. Effect of pore diameter on the phase behaviour of confined gases of hydrocarbon mixture under confined graphitic slits has been understood recently \cite{William2016}. Kazemi et al. \cite{Kazemi2017} very recently undertook a molecular simulation study on adsorption and transport of gases in carbon based organic nano-capillaries. Also some investigations have been reported on gas flow in confined spaces using both MD as well as continuum model \cite{Kiril2017,Volkan2015,Lei2016}.\\
 Svensson et al. \cite{Svensson2013} reported an experimental work of high resolution spectroscopy and assess pore size by studying wall collision broadening of absorption lines of gases in confinement. Granular gases under confinement have also gathered attention in the domain of confined gas studies. Florence et al. \cite{Florence2000} have done the experimental measurement of the spectrum of velocity fluctuations in a confined granular gas. Recently theoretical investigation on linear hydrodynamic stability analysis of confined granular gas has been performed \cite{Javier2016}.\\

Though many studies have been performed for confined gases using experimental, theoretical and MD methods, there is a lack of detailed and systematic findings related to the short-time dynamics, correlations and other features of atomistic origin through velocity autocorrelation function (VACF) and mean squared displacement (MSD) analyses of confined gases. As the transient caging phenomena, observed in liquids \cite{Alder1970,Hurley}, is not expected in dilute gaslike fluids due to its high diffusivity compared to the liquids, it is intriguing to ask if confinement influences the particle dynamics and if so, would it manifest in the VACF and MSD. \\

We present a study of molecular dynamics of gaslike fluid in partial confinement using LJ potential to address this question. The simulation domain is taken to be a cuboid with a pair of parallel sides to simulate partial confinement. The walls have been simulated as a smooth surface with reflective boundary condition, which generate force on the particles only in the normal direction. The separation between these reflective walls is varied from a few atomic diameters to large values. The velocity auto-correlation function (VACF) and the mean squared displacement (MSD) are studied as a function of the separation between the parallel reflective walls. We address the anisotropic features of VACF and MSD under confinement, observe correlations, occurred between them and establish the fact of spontaneous formation of sound-like waves by computing sound speed using VACF and MSD normal to the walls. The details of the MD simulation method is described in Section II .Section III contains the results and discussions followed by summary and conclusions in Section IV.

\section{\label{sec:2}Computational Details}

We carry out molecular dynamics calculations on LJ fluid using LAMMPS software package \cite{Pimpton1995}. We model $20000$ particles of LJ fluid fitted to argon properties (Mass of argon = $6.69 \times 10^{-26}$ kg) in gas phase both in bulk and in partially confined geometries. For bulk simulation, periodic boundary conditions (PBC) are imposed along each of the three dimensions at $300$K. Constant pressure temperature ensemble (NPT) is used to realize the dynamics of bulk argon gas for a wide range of pressures (from P = $0.004$ MPa to P = $3$ MPa) at $300$K. We impose the repulsive part of LJ-potential between the gas particles by setting the cut-off distance equal to the diameter ($\sigma$ = $3.4$ $\AA$) of LJ particles (argon).\\
After an energy minimization, standard velocity-verlet algorithm with a time-step ($\Delta t$) of $0.0001$ picosecond (ps) has been used for each of these systems (both bulk and confined) to equilibrate up to $10$ ns ($10^8$ steps) followed by a $2$ ns ($2\times 10^7$ steps) production run to calculate the properties of interests. During equilibration, the fluctuations of temperature, potential energy and kinetic energy have been monitored to ensure convergence. The temperature of the system has been controlled via Nose-Hoover thermostat.\\

In bulk simulations, equilibrium velocity distribution for each component ($v_x$,$v_y$,$v_z$) follows the Maxwell distribution corresponding to $300$K with the standard deviation of the distribution matching very well with the simulated data (analytical: $2.49$ $\AA/ps$, simulated: $2.51$ $\AA/ps$). For a test NPT simulation (P = $0.09$ MPa, T= $300$ K) the VACF is found to decay as exp($-t/\tau$), where $\tau$ is the relaxation time. We obtain $\gamma$ = $1/\tau$ from the fitted exponential as $0.0024$ $ps^{-1}$. Using $D$, the diffusion coefficient, by fitting the long-time average mean squared displacement (MSD), averaged over all the particles, to the Einstein’s relation $\left\langle\Delta r^2\right\rangle$ = $6Dt$ for a three dimensional system, obtained from simulation and estimate $\gamma$ from the analytical expression as $\gamma$ = $\alpha/m$ = $k_{B}T/mD$ = $0.0024$ $ps^{-1}$, prove the self-consistency of our MD results. For confinement studies, partially confined systems of LJ gas are simulated in a cuboid with $20000$ argon particles with reflecting parallel walls facing each other normal to the $z$ axis at $z$ = $\pm \frac{H}{2}$, $H$ being the separation between the walls. The walls are smooth and generate force on the particles only in the normal direction. The confinement studies have been carried out at $300$K for pressures ranging from $0.004$ MPa to $3$ MPa simulating a wide range of densities using NPT ensemble. It may be noted that above $5$ MPa argon enters supercritical regime at $300$K. For each of these (P,T) state points, the separation (H) between these reflective walls is varied from $20$ $\AA$ to $300$ $\AA$ ($6$$\sigma$ $\leq$ H $\leq$ $90$$\sigma$) with a step of $20$ $\AA$ ($\approx$ $6$ $\sigma$) such that the simulated density is same as that of the bulk for the corresponding (P,T) state point. Indeed, our MD simulated densities have been found to match quite well (relative error $\approx$ $1$ $\%$) to the experimentally measured density from the NIST database \cite{NIST}. Periodic boundary conditions are applied along $x$ and $y$ axes for all the partially confined systems.  

\section{\label{sec:3}Results and Discussions}

\subsection{\label{sec:level1}Effect of partial confinement on VACF:}
We chose a range of pressures (/densities) varying from higher to a very lower value at $300$K temperature of gaseous argon from NIST data \cite{NIST}. Constant pressure-temperature (NPT) ensembles have been used to realize the dynamics. The normalized VACF ($Z(t)$) is usually defined as 
\begin{equation}
Z(t) = \frac{\left\langle \sum_{j=1}^{N} \vec{v}_{j}(t) \vec{v}_{j}(0) \right\rangle}{\left\langle \sum_{j=1}^{N} \vec{v}_{j}(0) \vec{v}_{j}(0) \right\rangle}
\end{equation}
which contains the sum of the VACF along $x$,$y$ and $z$ directions, where $\vec{v}_{j}(0)$ and $\vec{v}_{j}(t)$ are velocity vectors of same particles (index $j$) at some initial $t$ = $0$ and at some later time ($t$) respectively and $\left\langle ... \right\rangle$ denotes the ensemble average. We use the notation of VACF$_\parallel$ and VACF$_\perp$ to designate the VACF along parallel ($x$, $y$) and perpendicular ($z$) directions with respect to the walls. The normalized versions of VACF$_\parallel$ ($Z_{xy}(t)$) and VACF$_\perp$ ($Z_{z}(t)$) are defined as
\begin{equation}
Z_{xy}(t) = \frac{\left\langle \sum_{j=1}^{N} \vec{v}_{x j}(t) \vec{v}_{x j}(0) \right\rangle + \left\langle \sum_{j=1}^{N} \vec{v}_{y j}(t) \vec{v}_{y j}(0) \right\rangle }{\left\langle \sum_{j=1}^{N} \vec{v}_{x j}(0) \vec{v}_{x j}(0) \right\rangle + \left\langle \sum_{j=1}^{N} \vec{v}_{y j}(0) \vec{v}_{y j}(0) \right\rangle}
\end{equation}
and
\begin{equation}
Z_{z}(t) = \frac{\left\langle \sum_{j=1}^{N} \vec{v}_{z j}(t) \vec{v}_{z j}(0) \right\rangle}{\left\langle \sum_{j=1}^{N} \vec{v}_{z j}(0) \vec{v}_{z j}(0) \right\rangle}
\end{equation}
\onecolumngrid
\begin{widetext}
\begin{center}
\begin{figure}[H]
\begin{minipage}{0.5\textwidth}
\includegraphics[width=1.0\textwidth]{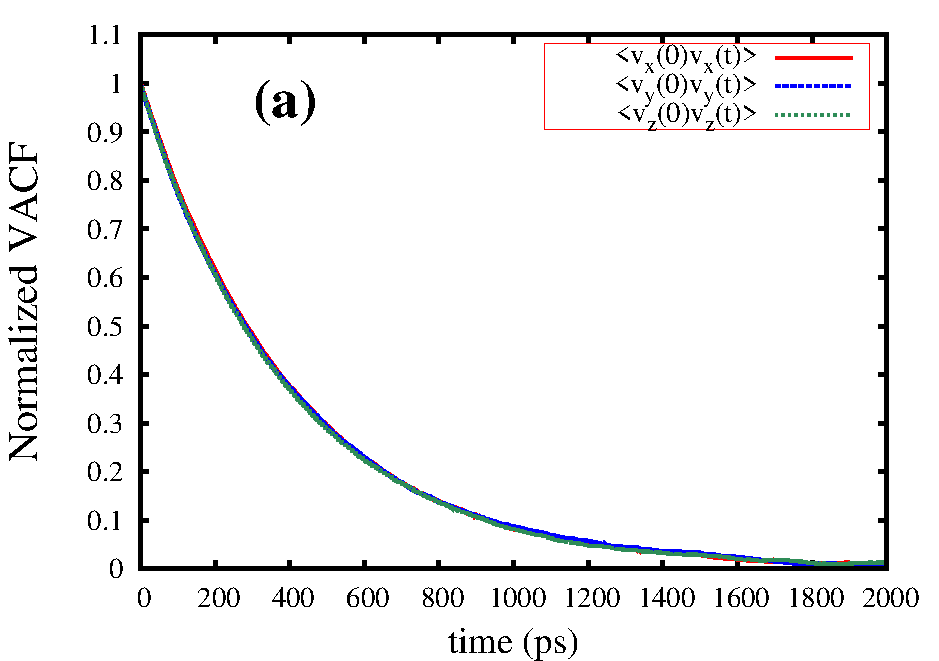}%
\end{minipage}
\begin{minipage}{0.5\textwidth}
\includegraphics[width=1.0\textwidth]{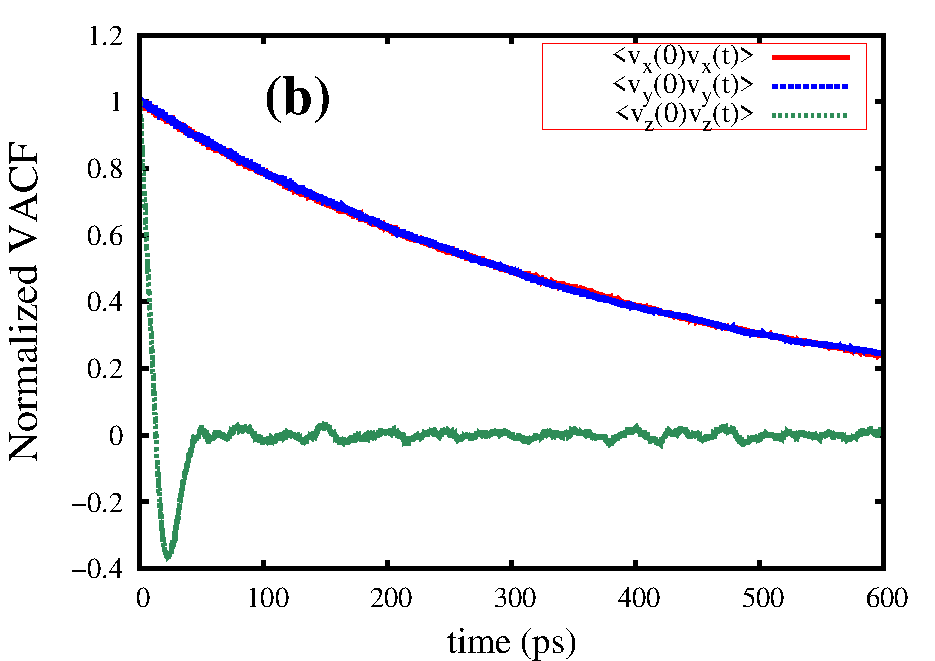}%
\end{minipage}
\caption{\label{1}\textbf{(a)} VACF in Bulk argon gas:Normalized VACF$_\parallel$(along $x$ and $y$ directions) and Normalized VACF$_\perp$($z$)(along $z$ direction) follow identical monotonic decay trend, simulated under constant pressure and temperature of $300$K and $0.09$MPa respectively. \textbf{(b)} Difference between Normalized VACF$_\parallel$(along $x$ and $y$) and Normalized VACF$_\perp$(along $z$) under partial  confinement: Simulations mimic argon gas with P = $0.09$ MPa, T = $300$ K and with wall spacing of $100$ $\AA$ ($\approx$ $30$ $\sigma$). When compared to the trends shown by VACF$_\parallel$, the VACF$_\perp$ features rapid decay suggesting that confinement suppresses the role of thermal fluctuations significantly.}
\end{figure}
\end{center}
\vspace{-1cm}
\end{widetext}
Here, $\vec v_{xj}(0)$, $\vec v_{yj}(0)$, $\vec v_{zj}(0)$ and $\vec v_{xj}(t)$, $\vec v_{yj}(t)$, $\vec v_{zj}(t)$ denote velocities of $j^{th}$ particle along $x$, $y$ and $z$ directions at initial and at some later time $t$ respectively, $N$ is the total number of particles and $\left\langle ... \right\rangle$ denotes the ensemble average. For the bulk case (Fig.\ref{1}.(a)), no oscillatory behaviour is observed, and both VACF$_\parallel$ and VACF$_\perp$ follow a monotonous decay trend. It validates the well known fact that the VACF is a monotonically decreasing function in the bulk gaseous environment but it shows oscillating nature in liquid and solid phases \cite{Krishnan2003}.\\

In Fig.\ref{1}.(b) we observe that in partially confined gaseous argon the VACF$_\parallel$ shows identical monotonous decay as that of the bulk. As the system is "infinite" along $x$ and $y$ axes (due to PBC) the monotonic decay of VACF$_\parallel$ appears to be due to particle-particle collisions as in the bulk. As may be expected, the presence of the walls has no effect on VACF$_\parallel$.\\

In sharp contrast to the trends shown by VACF$_\parallel$, VACF$_\perp$ shows a prominent minimum at short time scale followed by a flat plateau like regime (Fig.\ref{1}.(b)). The early onset of the plateau like regime is due to the large number of collisions with the walls. To ensure that the VACF-minima in our confined cases normal to the wall direction are not influenced by the artifacts coming from $x$ and $y$ directions through periodic boundaries, we monitored the VACF$_\perp$ at different time intervals starting from the beginning of our simulation and found the minima of VACF$_\perp$ are occurring at precisely the same time for a fixed wall separation.\\

\subsection{\label{sec:level1}Effect of partial confinement on MSD:}
The ensemble-averaged Mean-squared displacement is defined as:
\begin{equation}
MSD(t) = \frac{1}{N}\left\langle \sum_{j=1}^N\left[\vec{R}_{j}(t)-\vec{R}_{j}(0)\right]^2 \right\rangle
\end{equation}
 where, $\vec{R}_{j}(0)$ and $\vec{R}_{j}(t)$ are the position vectors of same particles (index $j$) at some initial $t$ = $0$ and at some later time ($t$) respectively, N is the total number of particles and $\left\langle ... \right\rangle$ denotes the ensemble average. We study both MSD$_\parallel$ (MSD along $x$ and $y$ directions) and MSD$_\perp$ (MSD along $z$ direction) for bulk as well as for partially confined gas as a function of time. MSD$_\parallel$ and MSD$_\perp$ are defined as
\begin{equation}
MSD_{\parallel}(t) = \frac{1}{N}\left[\left\langle \sum_{j=1}^N\left[x_{j}(t)-x_{j}(0)\right]^2 \right\rangle + \left\langle \sum_{j=1}^N\left[y_{j}(t)-y_{j}(0)\right]^2 \right\rangle\right]
\end{equation}
and
\begin{equation}
MSD_{\perp}(t) = \frac{1}{N}\left\langle \sum_{j=1}^N\left[z_{j}(t)-z_{j}(0)\right]^2 \right\rangle
\end{equation}
where, $x_j(0)$, $y_j(0)$, $z_j(0)$ and $x_j(t)$, $y_j(t)$, $z_j(t)$ are positions of same particles (index $j$) along $x$, $y$ and $z$ at initial and at some later time $t$ respectively with $\left\langle ... \right\rangle$ denoting the ensemble average. In the bulk case (Fig.\ref{2}.(a)), we observe an identical trend for both MSD$_\parallel$ (MSD along $x$ and $y$) and MSD$_\perp$ (MSD along $z$) as a function of time, with an initial short-time ballistic (MSD $\sim$ $t^2$), followed by a long-time diffusive motion (MSD $\sim$ $t$).
\onecolumngrid
\begin{widetext}
\begin{center}
\begin{figure}[H]
\begin{minipage}{0.5\textwidth}
\includegraphics[width=1.0\textwidth]{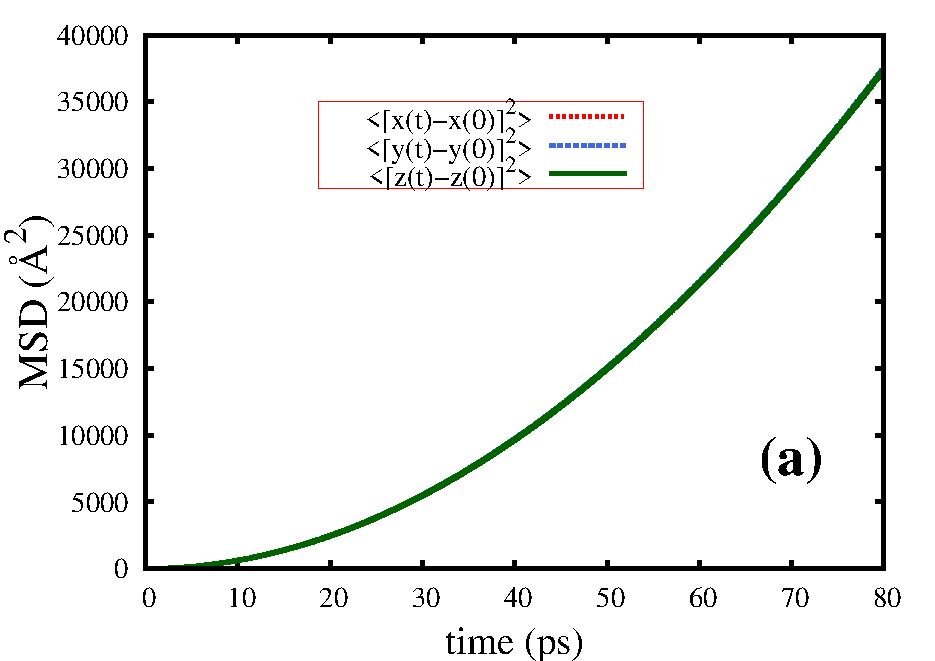}%
\end{minipage}
\begin{minipage}{0.5\textwidth}
\includegraphics[width=1.0\textwidth]{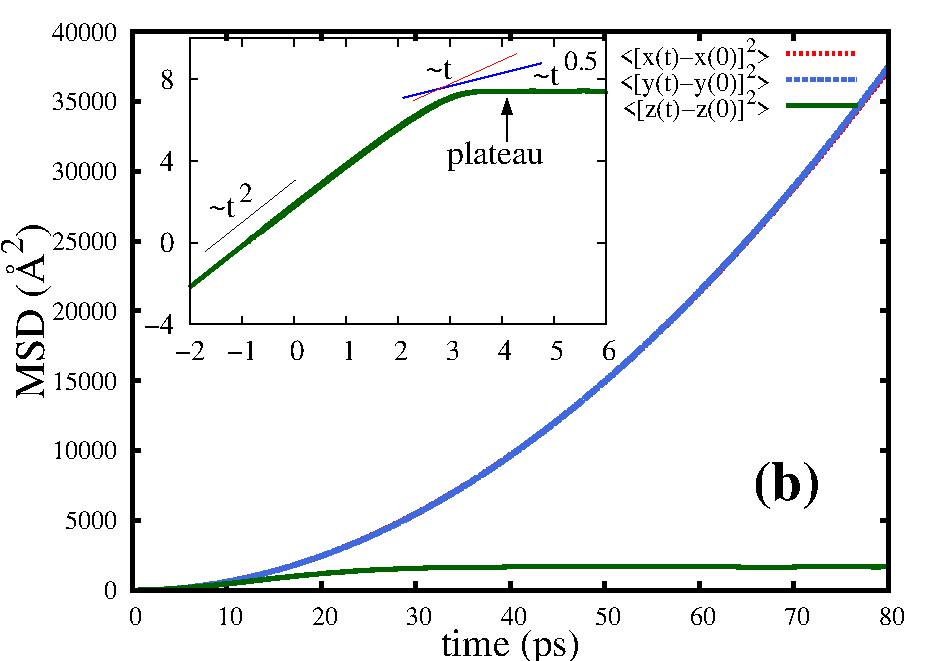}%
\end{minipage}
\caption{\label{2}\textbf{(a)} Isotropic nature of MSD in Bulk argon-like gas:MSD$_\parallel$(MSD along $x$ and $y$ directions) and MSD$_\perp$(MSD along $z$) follow identical trend, simulated under constant pressure and temperature of $300$K and $0.09$MPa respectively. \textbf{(b)} Comparison between MSD$_\parallel$($x$ and $y$) and MSD$_\perp$($z$) under partial  confinement: MSD$_\perp$ deviates from MSD$_\parallel$ exhibiting sub-diffusive behavior before reaching a plateau: Simulations mimic argon-like gas with P = $0.09$ MPa, T = $300$ K and with wall spacing of $100$ $\AA$ ($\approx$ $30$ $\sigma$). The inset shows different regimes of MSD$_\perp$ ($MSD_{Z}$) with different slopes as a function of time in a log-log scale. The gradual crossover of MSD$_\perp$ from ballistic ($\sim$ $t^2$) to diffusive ($\sim$ $t$) to sub-diffusive ($\sim$ $t^{0.5}$) is shown for P = $0.09$ MPa, T = $300$ K and with wall spacing of $100$ $\AA$ ($\approx$ $30$ $\sigma$).}
\end{figure}
\end{center}
\vspace{-1cm}
\end{widetext}
The isotropic nature of MSD is evident. The diffusion coefficient can be extracted from the long-time MSD using $\left\langle\left(\Delta r\right)^2\right\rangle$ = $2nD\Delta t$, where $n$ stands for dimensions involved in the system and $D$: the diffusion coefficient.\\
 
In the partially confined case shown in Fig.\ref{2}.(b), MSD$_\perp$ shows three distinct regimes: ballistic-like motion on short time scales (MSD$_\perp$ scales quadratically with time), diffusive motion over intermediate time scales (MSD$_\perp$ scales linearly with time) and anomalous diffusion over long time scales (MSD$_\perp$ scales non-linearly with time). Sub-diffusive motion refers to such anomalous diffusion where MSD$_\perp$ $\sim$ $t^\alpha$, where $0<\alpha <1$. The inset in Fig.\ref{2}.(b) shows clearly the transition from diffusive to sub-diffusive motion. Over long timescales the frequent collisions between the walls and the other particles impose severe constraints on the MSD$_\perp$ to give rise to a constant plateau.  
\onecolumngrid 
\begin{widetext}
\begin{center}
\begin{figure}[H]
\begin{minipage}{0.5\textwidth}
\includegraphics[width=1.0\textwidth]{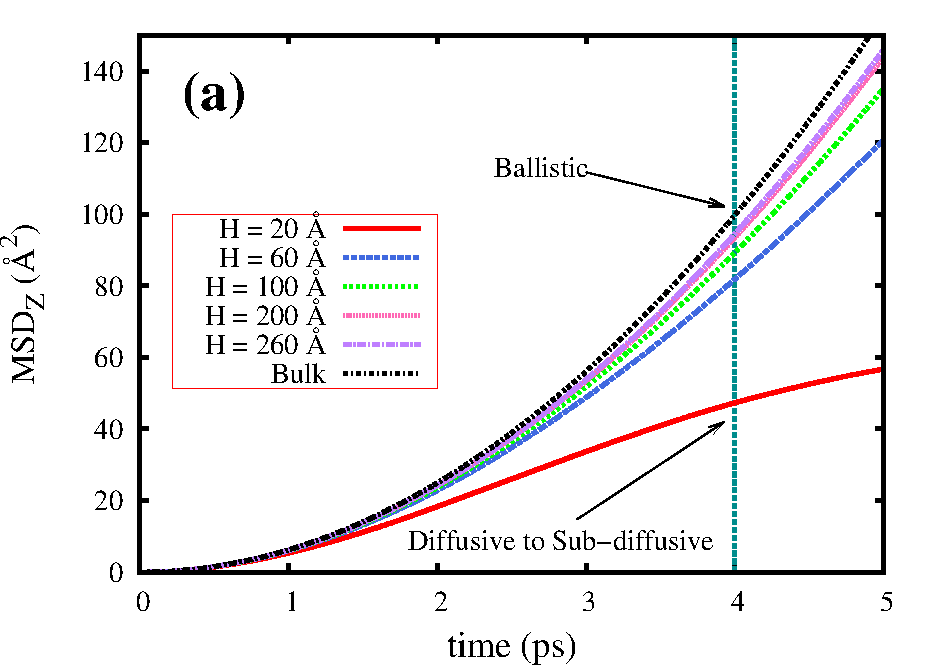}%
\end{minipage}
\begin{minipage}{0.5\textwidth}
\includegraphics[width=1.0\textwidth]{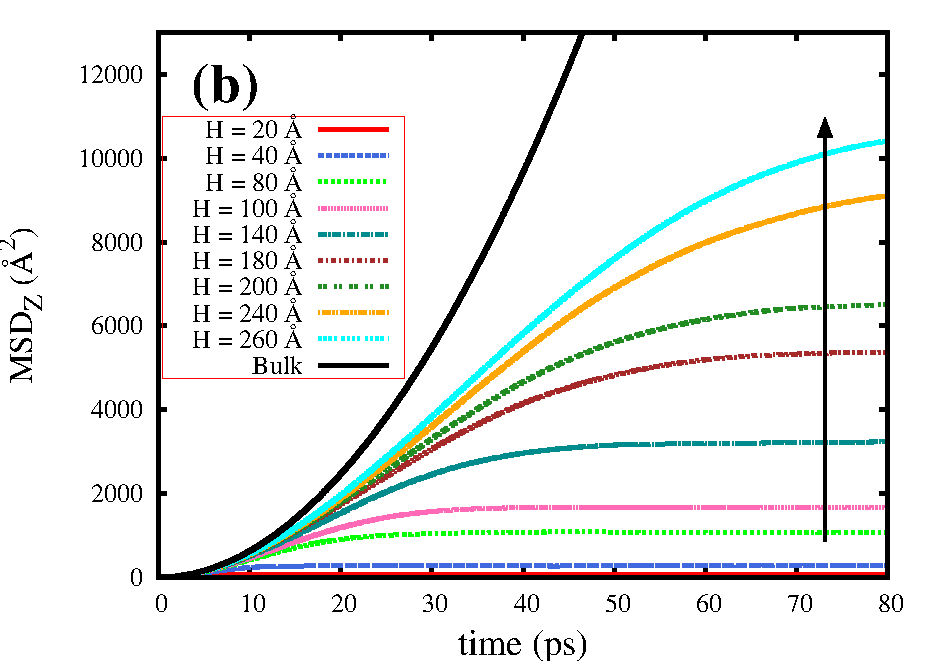}%
\end{minipage}
\caption{\label{3}\textbf{(a)} Short-time MSD$_\perp$ ($MSD_{Z}$) with different confined spacings for argon gas at P=$0.09$MPa and T=$300$K. At $t$ = $4$ ps (shown by a vertical line) the confined system (H = $20$ $\AA$ $\sim$ $6$ $\sigma$) shows diffusive to sub-diffusive transition while the bulk MSD$_\perp$ still shows ballistic-like motion. \textbf{(b)} Long-time MSD$_\perp$ ($MSD_{Z}$) with different confined spacings (H) for argon gas at P=$0.09$MPa and T=$300$K. The arrow indicates the direction of increasing H. When compared to the trends shown by MSD$_\parallel$, the MSD$_\perp$ features subdiffusive behavior,
suggesting that confinement suppresses the role of thermal fluctuations significantly.}
\end{figure}
\end{center}
\end{widetext}

\begin{table}[H]
\centering
\caption{\label{table:1} Variation of time-duration and proportionality constant $a$ in the ballistic regime of argon with P = $0.09$MPa and T = $300$K for different confined spacings. For reference the bulk value is also included.}
\begin{tabular}{|p{2.1cm}||p{2.0cm}||p{1.5cm}||p{2.4cm}|} 
\hline
& & &\\
\textbf{H (spacing)} & \textbf{time duration of ballistic motion (ps)} & \textbf{a(\textbf{$\AA^2/ps^2$}) from MD} & \textbf{\textbf{$k_{B}T/M$}(\textbf{$\AA^2/ps^2$}) Analytical value} \\ [0.5ex]
\hline
$20$ $\AA$ ($\approx$ $6$ $\sigma$) & $0.7$ & $5.788$ &  \\
$60$ $\AA$ ($\approx$ $18$ $\sigma$) & $1.92$ & $5.933$ &  \\
$100$ $\AA$ ($\approx$ $30$ $\sigma$) & $3.7$ & $5.723$ & $6.191$ \\
$200$ $\AA$ ($\approx$ $59$ $\sigma$) & $5$ & $5.908$ & \\
$260$ $\AA$ ($\approx$ $76$ $\sigma$) & $6$ & $5.867$ & \\
Bulk & $95$ & $5.870$ & \\[1ex] 
\hline
\end{tabular}
\end{table}    
It can also be seen from Fig.\ref{2}.(b) that MSD$_\parallel$ is barely affected by the partial confinement.

\subsection{\label{sec:level1} Effect of varying confined spacing (H) on $MSD_\perp$:}

We study the variation of the mean squared displacement along $z$ (MSD$_\perp$) as a function of time for a range of wall spacings. The trends in MSD$_\perp$ from Fig.\ref{3}.(a) and (b) indicate that the temporal windows in which ballistic, diffusive, and sub-diffusive motion prevail are strongly influenced by H, the wall spacing. Smaller the H, shorter the time windows are.\\

For comparison, the bulk MSD$_\perp$ is also included in Fig.\ref{3}.(a). While in the bulk, ballistic motion, characterized by a $t^2$ dependence in MSD$_\perp$ (MSD$_\perp$ $\sim$ $at^2$), can be seen to exist upto $95$ ps, it lasts only up to $0.7$ ps for the smallest confinement spacing (H = $20$ $\AA$ $\approx$ $6$ $\sigma$) (Table.\ref{table:1}). \\

It is also observed from Table.\ref{table:1} that irrespective of the spacing, the proportionality factor, $a$, remains almost constant and close to that of the bulk. Likewise, the diffusion coefficients ($D_{\perp}$), evaluated within the time window marked by the onset of diffusive motion and by the transition from diffusive to sub-diffusive motion in MSD$_\perp$, reduce systematically as H is reduced as can be seen from Table.\ref{table:2}. \\

Further, the MSD$_\perp$ undergoes a transition from diffusive to sub-diffusive behavior which correlates well with the wall spacing as can be seen qualitatively from Fig.\ref{2}.(b) and Fig.\ref{3}.(b), and quantitatively, from Table.\ref{table:3}. \\

Columns $2$, $4$, $6$, and $8$ of Table.\ref{table:3} present the average root mean-squared displacement of the particles during 
\begin{table}[H]
\centering
\caption{\label{table:2} Variation of diffusion coefficients ($D_{\perp}$) evaluated from the diffusive regime in MSD$_\perp$ as a function of H for highest ($3$ MPa) and lowest pressure ($0.004$ MPa) studied. For reference the corresponding bulk values are also included.}
\begin{tabular}{|p{2.1cm}||p{2.5cm}||p{2.6cm}|} 
\hline
& &\\
\textbf{P (MPa)} & \textbf{H (spacing)} & \textbf{$D_{\perp}$ (\textbf{$\AA^2/ps$}) from MSD$_\perp$} \\ [0.5ex]
\hline
& & \\
$0.004$ & $20$ $\AA$ ($\approx$ $6$ $\sigma$) & $7.73$  \\
& $60$ $\AA$ ($\approx$ $18$ $\sigma$) & $22.99$  \\
& $100$ $\AA$ ($\approx$ $30$ $\sigma$) & $37.89$ \\
& $200$ $\AA$ ($\approx$ $59$ $\sigma$) & $78.17$ \\
& $300$ $\AA$ ($\approx$ $90$ $\sigma$) & $118.03$ \\
& Bulk & $565.69 \times 10^3$ \\
\hline
& & \\
$3$ & $20$ $\AA$ ($\approx$ $6$ $\sigma$) & $7.15$  \\
& $60$ $\AA$ ($\approx$ $18$ $\sigma$) & $18.27$  \\
& $100$ $\AA$ ($\approx$ $30$ $\sigma$) & $27.49$ \\
& $200$ $\AA$ ($\approx$ $59$ $\sigma$) & $41.03$ \\
& $300$ $\AA$ ($\approx$ $90$ $\sigma$) & $48.82$ \\
& Bulk & $77.29$ \\[1ex] 
\hline
\end{tabular}
\end{table}
the diffusive-sub-diffusive transition, $\sqrt{MSD_{\perp}(diff)}$, for four different pressures. Except for the highest pressure considered, this transition can be seen to occur when particles traverse an average distance $\sim$ $H/3$, where H is the wall spacing. At the highest pressure, the average distance traversed is consistently smaller ($\sim$ $H/3.9$). Columns 3, 5, 7, and 9 of Table.\ref{table:3} present the asymptotic average root mean squared displacement of the particles, $\sqrt{MSD_{\perp}(sat)}$, for the four different pressures. We see that for all the P, T state points and for all the spacings considered, the values for $\sqrt{MSD_{\perp}(sat)}$ are nearly constant ($\sim$ $H/2.45$) indicating that the particles traverse this distance and are constrained to be no closer than this distance on the average.\\

It is clear from the observed trends in MSD$_\perp$ that confinement alters the purely thermal inter-particle collisional characteristics over two time scales in different ways. We note that the confinement results in two new processes: a particle undergoes collisions with other particles that bounces-off from the walls and termed as wall-mediated collisions hereinafter and a particle undergoes direct collisions with the walls. Wall-mediated collisions occur over short and intermediate time scales and the direct particle-wall collisions occur over intermediate and long time scales. Both of these collisions are non-thermal in nature and modify the thermal collisions to varying degrees depending on the wall spacing.\\

In the ballistic regime, non-thermal inter-particle 
\onecolumngrid
\begin{widetext}
\begin{table}[H]
\centering
\caption{\label{table:3} Variation of $H/\sqrt{MSD_{\perp}(diff)}$ and $H/\sqrt{MSD_{\perp}(sat)}$ of argon gas as a function of confined spacing (H) for four different pressures at T = $300$ K.}
\begin{adjustbox}{max width=1.07\textwidth}
\begin{tabular}{|c|c|c|c|c|c|c|c|c|} 
\hline 
& \multicolumn{2}{|c|}{\textbf{P = $0.004$ MPa}} & \multicolumn{2}{|c|}{\textbf{P = $0.03$ MPa}} & \multicolumn{2}{|c|}{\textbf{P = $0.3$ MPa}} & \multicolumn{2}{|c|}{\textbf{P = $3$ MPa}} \\
& & & & & & & &\\
\textbf{H}& \textbf{H/$\sqrt{MSD_{\perp}(diff)}$} & \textbf{H/$\sqrt{MSD_{\perp}(sat)}$} & \textbf{H/$\sqrt{MSD_{\perp}(diff)}$} & \textbf{H/$\sqrt{MSD_{\perp}(sat)}$} & \textbf{H/$\sqrt{MSD_{\perp}(diff)}$} & \textbf{H/$\sqrt{MSD_{\perp}(sat)}$} & \textbf{H/$\sqrt{MSD_{\perp}(diff)}$} & \textbf{H/$\sqrt{MSD_{\perp}(sat)}$} \\ [0.5ex]
\hline
& & & & & & & &\\
$20$ $\AA$ ($\approx$ $6$ $\sigma$) & $3.04$ & $2.45$ & $2.97$  & $2.45$ & $3.04$ & $2.46$ &  $3.11$ & $2.44$ \\
& & & & & & & &\\
$40$ $\AA$ ($\approx$ $12$ $\sigma$) & $3.07$ & $2.45$ & $3.05$  & $2.46$ & $3.09$ & $2.45$ & $3.19$ & $2.44$ \\
& & & & & & & &\\
$60$ $\AA$ ($\approx$ $18$ $\sigma$) & $3.05$ & $2.45$ & $3.08$ & $2.44$ & $3.10$ & $2.45$ & $3.48$ & $2.45$ \\
& & & & & & & &\\
$80$ $\AA$ ($\approx$ $24$ $\sigma$) & $3.08$ & $2.45$ & $3.05$  & $2.47$ & $3.15$ & $2.44$ & $3.39$ & $2.45$ \\
& & & & & & & &\\
$100$ $\AA$ ($\approx$ $30$ $\sigma$) & $3.02$ & $2.45$ & $3.04$  & $2.46$ & $3.16$ & $2.45$ & $3.69$ & $2.44$ \\
& & & & & & & &\\
$120$ $\AA$ ($\approx$ $36$ $\sigma$) & $3.07$ & $2.46$ & $3.09$  & $2.46$ & $3.17$ & $2.45$ & $3.75$ & $2.45$ \\
& & & & & & & &\\
$140$ $\AA$ ($\approx$ $42$ $\sigma$) & $3.06$ & $2.47$ & $3.05$  & $2.46$ & $3.06$ & $2.45$ & $3.88$ & $2.45$ \\
& & & & & & & &\\
$160$ $\AA$ ($\approx$ $47$ $\sigma$) & $3.16$ & $2.46$ & $3.15$  & $2.47$ & $3.19$ & $2.46$ & $3.86$ & $2.45$ \\
& & & & & & & &\\
$180$ $\AA$ ($\approx$ $53$ $\sigma$) & $3.13$ & $2.45$ & $2.98$  & $2.46$ & $3.25$ & $2.45$ & $3.99$ & $2.45$ \\
& & & & & & & &\\
$200$ $\AA$ ($\approx$ $59$ $\sigma$) & $3.09$ & $2.46$ & $3.06$  & $2.48$ & $3.28$ & $2.45$ & $4.22$ & $2.45$ \\
& & & & & & & &\\
$220$ $\AA$ ($\approx$ $65$ $\sigma$) & $3.08$ & $2.44$ & $3.06$  & $2.45$ & $3.19$ & $2.46$ & $4.20$ & $2.45$ \\
& & & & & & & &\\
$240$ $\AA$ ($\approx$ $71$ $\sigma$) & $3.04$ & $2.44$ & $3.08$  & $2.46$ & $3.26$ & $2.45$ & $4.37$ & $2.44$ \\
& & & & & & & &\\
$260$ $\AA$ ($\approx$ $76$ $\sigma$) & $3.04$ & $2.45$ & $3.09$  & $2.45$ & $3.17$ & $2.45$ & $4.40$ & $2.45$ \\
& & & & & & & &\\
$280$ $\AA$ ($\approx$ $82$ $\sigma$) & $3.14$ & $2.45$ & $3.09$  & $2.45$ & $3.23$ & $2.45$ & $4.21$ & $2.45$ \\
& & & & & & & &\\
$300$ $\AA$ ($\approx$ $90$ $\sigma$) & $3.12$ & $2.46$ & $3.07$  & $2.45$ & $3.23$ & $2.44$ & $4.27$ & $2.45$ \\ [1ex] 
\hline
\end{tabular}
\end{adjustbox}
\end{table}
\end{widetext}
collisions are ineffective as the value of $a$ is close to that in the bulk. In the diffusive regime of MSD$_\perp$, the non-thermal  collisions reduce the diffusion coefficient ($D_{\perp}$) significantly from that in the bulk. The diffusive to sub-diffusive transition, and the constraint on MSD$_\perp$ not to exceed a characteristic length ($\sim$ $H/2.45$) in the sub-diffusive regime, are drastic manifestations of the non-thermal collisions resulting from the confinement.

\subsection{\label{sec:level1}The time-window between ballistic to diffusive transition:}
\begin{figure}[H]
\centering
\includegraphics[width=0.5\textwidth]{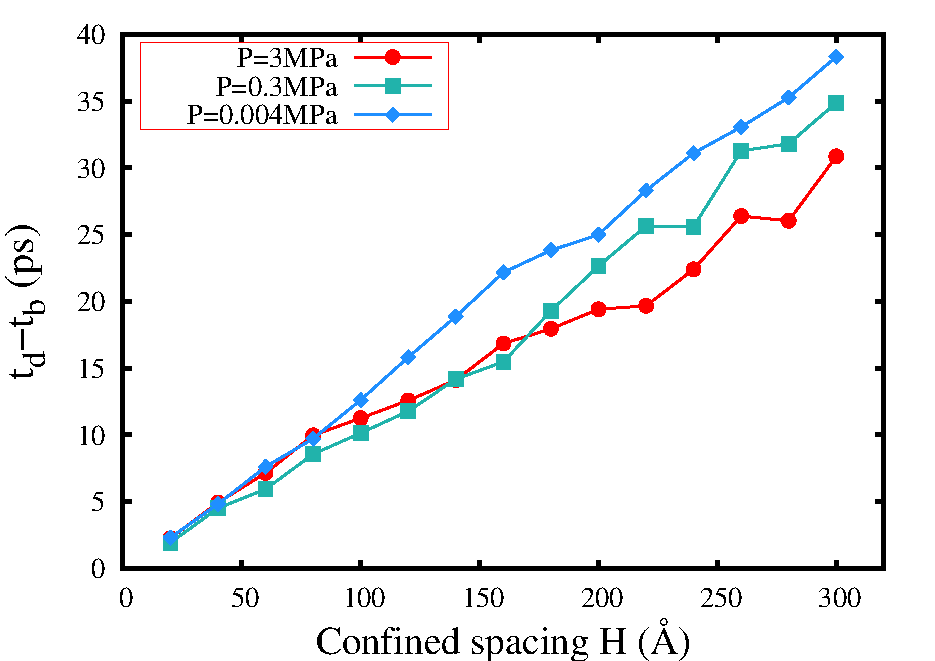}%
\caption{\label{4} Variation of time-window between ballistic to diffusive motion ($t_d$-$t_b$) as a function of confined spacing for different pressures.}
\end{figure}

Analyzing MSD$_\perp$ in greater detail, through Fig.\ref{2}.(b) in logarithmic scale, we observe that the particles undergo a gradual transition from ballistic to diffusive motion indicated by the slope changing gradually from $2$ to $1$. Fig.\ref{4} shows that the gradual transition occurring within a time-window between the end of ballistic motion ($t_b$) and the starting of diffusive motion ($t_d$) is found to be a function of wall spacing and pressure. The time window is found to be increasing with spacing for the whole range of pressures under study. Confnement enhances wall mediated collisional events much like the effect of increasing pressure except that the latter affects all the three degrees of freedom equally. In Fig.\ref{4}, three distinct regimes are observed. At very small wall spacings, values are independent of the pressure as the effect of confinement is strong.\\
 
At larger spacings, the pressure dependence prevails and is expected to be retained thereafter for greater spacings and asymptotically approach the corresponding bulk phase values. It is worth noting that, in the bulk gas, the time
window is largest ($1450$ ps) for the lowest fluid density (at the lowest pressure of $0.004$ MPa) considered and smallest ($229.67$ ps) for a relatively higher fluid density (at the highest pressure of $3$ MPa). \\

Over intermediate wall spacings, however, the crossover-like feature seen between wall spacings $H$ = $100$ $\AA$ ($\sim$ $30$ $\sigma$) and $H$ = $200$ $\AA$ ($\sim$ $59$ $\sigma$) in the figure suggests that there is a competition between the effects of confinement and the effects of pressure on the particle motion.\\

It may be noted that less than a decade ago, the full transition from ballistic to diffusive motion of a brownian particle in a liquid was observed experimentally \cite{Huang}. More specifically, by experimentally measured instantaneous velocity of Brownian particles in air, using optical tweezer, Tongcang Li et al. \cite{Tongcang} showed  the presence of a time-window between ballistic to diffusive motion of Brownian particles, although it must be noted that the timescales of Brownian motion are very different than that of self-diffusion in  the dilute gas discussed in this work.

\subsection{\label{sec:level1} Effect of varying confined spacing (H) on $VACF_\perp$ and correlation between $VACF_{\perp}-minima$ and MSD$_\perp$:}

\begin{figure}[H]
\centering
\includegraphics[width=0.5\textwidth]{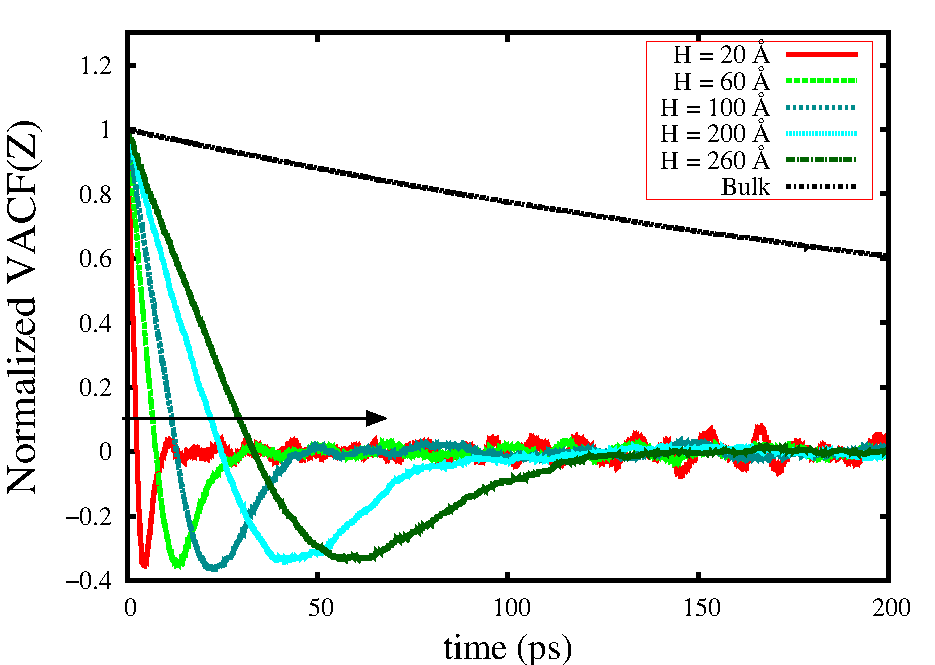}%
\caption{\label{5} Variation of VACF$_\perp$ with time for different confined spacings for argon gas at P=$0.09$MPa and T=$300$K. The arrow indicates the direction of increasing H.}
\end{figure}

Fig.\ref{5} presents the trends in the VACF$_\perp$ as a function of wall spacing at the same P, T state point considered when discussing the results for MSD$_\perp$ shown in Fig.\ref{3}.(b). As expected, VACF$_\parallel$ (not shown) is unaffected by confinement.\\
While the ballistic regime is barely visible on the time  scales shown in Fig.\ref{5}, the decay in the diffusive regime,  the occurrence of minima in the diffusive to sub-diffusive transition zone, and the approach to zero in the sub-diffusive regimes can be seen clearly. The initial decay in VACF$_\perp$, more rapid than what would result from purely thermal collisions, arises due to the wall-mediated collisions which are non-thermal in nature. As the spacing is reduced, the decay is faster indicating that wall-mediated collisions grow in importance at smaller spacings.\\
\begin{widetext}
\vspace{-0.2cm}
\begin{center}
\begin{figure}[H]
\begin{minipage}{.5\textwidth}
\includegraphics[width=1.0\textwidth]{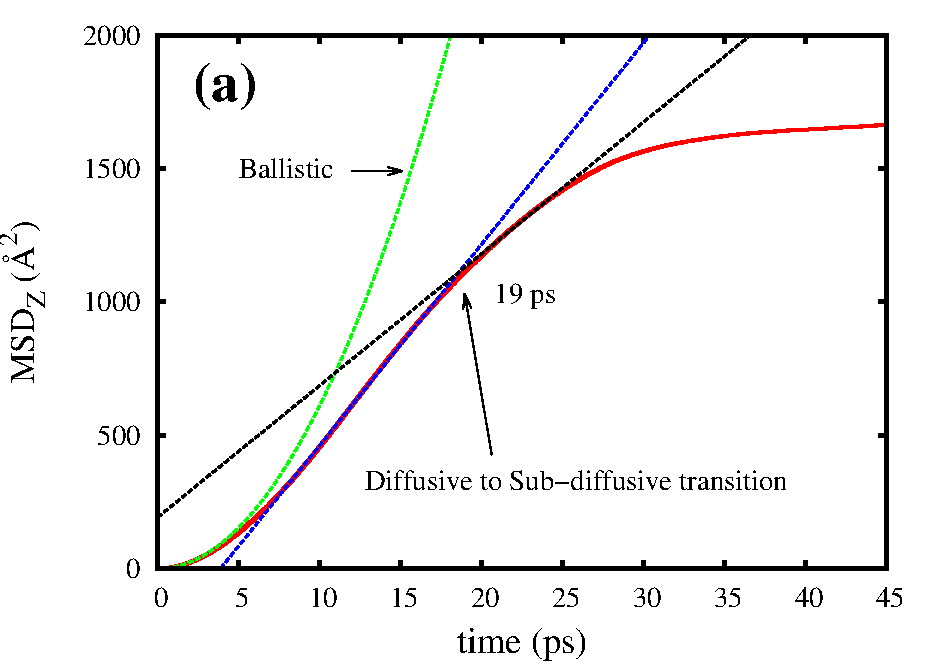}
\end{minipage}
\begin{minipage}{.5\textwidth}
\includegraphics[width=1.0\textwidth]{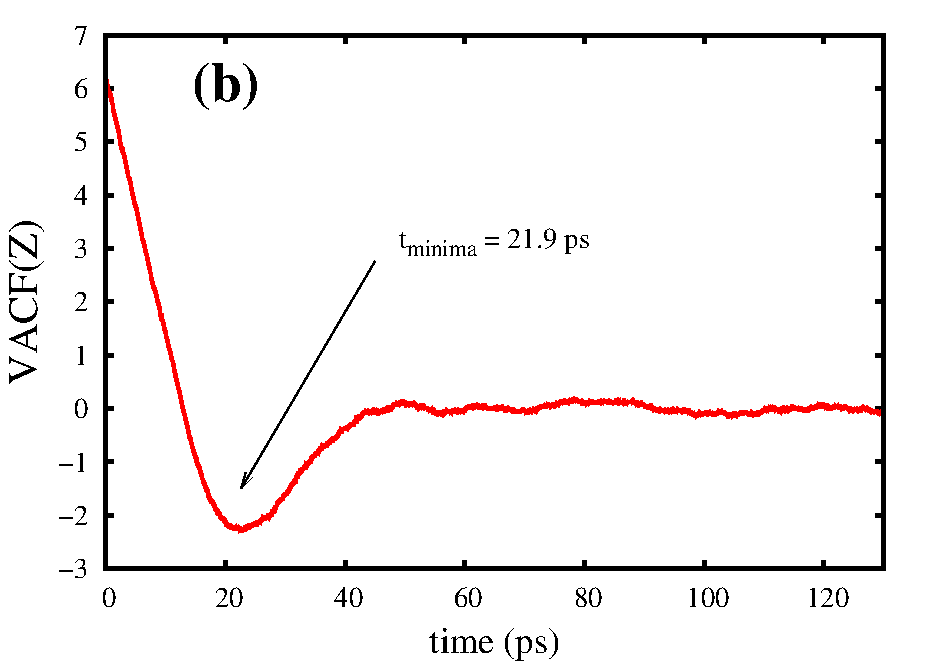}
\end{minipage}
\caption{\label{6} The transition time for MSD$_\perp$ ($MSD_{Z}$) to change from diffusive to sub-diffusive scaling corresponds to the minima occurring in VACF$_\perp$, shown for argon with T = $300$K and P = $0.09$ MPa. Fig \ref{6}.\textbf{(a)} shows $t_{tr}$ = $19$ ps in MSD$_\perp$, whereas Fig \ref{6}.\textbf{(b)} shows $t_{minima}$ = $21.9$ ps in VACF$_\perp$ for $100$ $\AA$ ($\approx$ $30$ $\sigma$) spacing.}
\end{figure}
\end{center}
\vspace{-1cm}
\end{widetext}

For any given wall spacing, as the particle approaches the walls over long time scales, the encounters with other particles bouncing-off the walls would increase and compete with direct particle-wall collisions. The resulting sub-diffusive motion manifests as a slow evolution in MSD$_\perp$ and a nearly vanishing VACF$_\perp$.\\

On intermediate time scales, the VACF$_\perp$ exhibits minima much like what is found in liquids.
The minima in VACF$_\perp$ can be seen to progressively shift to larger values in time as the wall-spacing is increased. It may also be noted that the depth of the minima is nearly the same for all the wall-spacings considered. However, these minima broaden gradually as wall-spacing increases. We observe a strong correlation between the diffusive to sub-diffusive transition in MSD$_\perp$ and the negative correlation minimum in the VACF$_\perp$­ for all partial confinements studied. Figure \ref{6} shows that the transition time ($t_{tr}$) from diffusive to sub-diffusive regimes in MSD$_\perp$, obtained graphically, is close to the time at which the minimum in VACF$_\perp$ occurs. For gaseous regime of argon with T = $300$ K and P = $0.09$ MPa, it is observed that $t_{tr}$ = $19$ ps, whereas $t_{min}$ is $21.9$ ps for a spacing of $100$ $\AA$ ($\approx$ $30$ $\sigma$).

\subsection{\label{sec:level1} Wall-mediated collisions and VACF$_\perp$-minima}

The minima in VACF$_\perp$, over intermediate time scales, arise due to the negative correlations produced by the velocity reversals from wall-mediated collisions. Whether a particle has a positive or a negative initial velocity in the $z$-direction, purely thermal collisions in the gas phase tend to decelerate (through frequent "soft" collisions) rather than produce direction reversals (through infrequent "hard" collisions). Under confinement, however, particles bouncing-off the walls are always moving away from the walls in the opposite direction inducing velocity reversals. These velocity reversals lead to negative correlations in the VACF$_\perp$ seen over intermediate time scales. As the wall spacing is reduced, non-thermal collisions dominate and the velocity reversals occur over shorter time scales leading to the early onset of the minimum in the VACF$_\perp$. For the larger spacings, the shift of minima to later times and the broadening of minima, observed in Fig.\ref{5}, indicate that thermal collisions mitigate the effects of non-thermal collisions only to a limited extent. \\

To understand the relative roles of thermal and non-thermal collisions, the VACF$_\perp$ is examined keeping the spacing fixed at H = $20$ $\AA$ ($\approx$ $6$ $\sigma$) and diluting the gas density from $4.81$ $kg/m^{3}$ (corresponding to P = $0.3$ MPa of argon gas at T = $300$ K) to a gas density of $0.012$ $kg/m^3$.   \\

From the results shown in Fig.\ref{7}, we can observe four  
\onecolumngrid
\begin{widetext}
\vspace{-0.2cm}
\begin{figure}[H]
\centering
\includegraphics[width=1.0\textwidth]{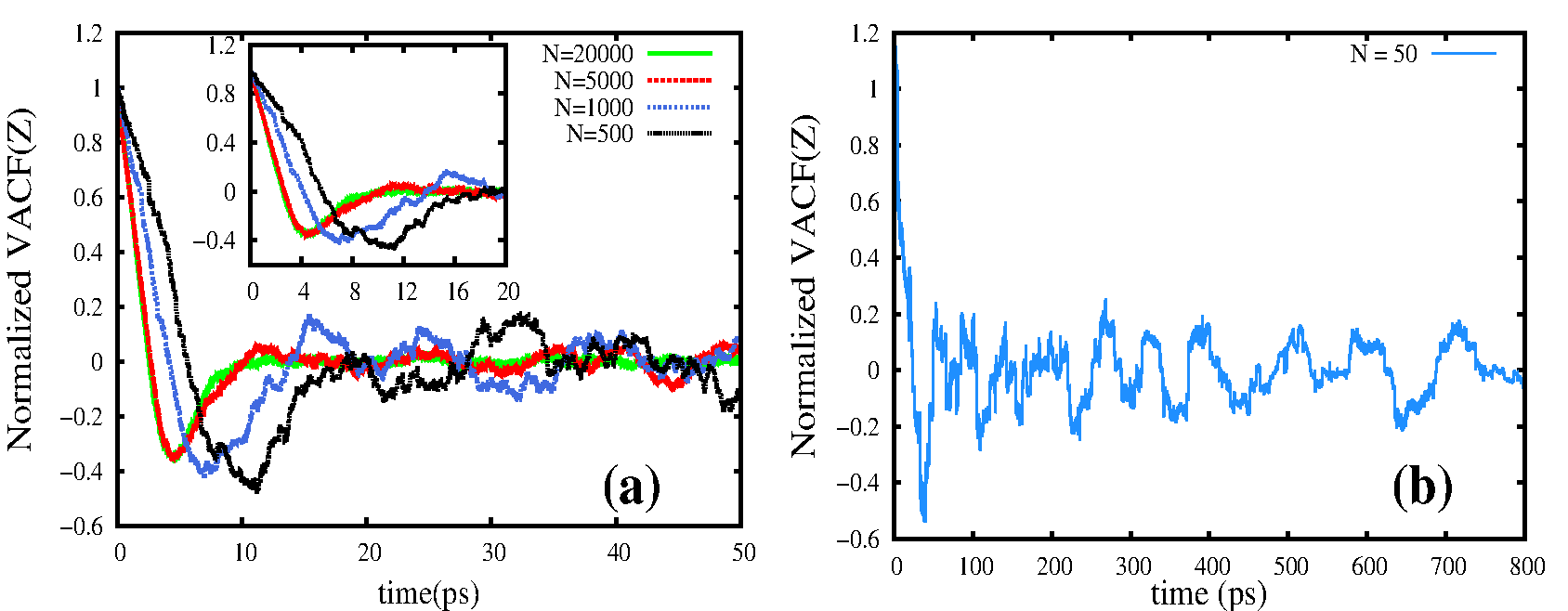}%
\caption{\label{7} (a).Normalized VACF$_\perp$ for different number densities of argon gas under same confined spacing (H = $20$ $\AA$ $\approx$ $6$ $\sigma$) at $300$ K. Distinct broadening of the first minimum and a systematic shift of the VACF-minima to later times are shown (see inset). (b) The oscillatory feature of VACF is seen at later times for very dilute argon gas (N = $50$) for H = $20$ $\AA$ ($\approx$ $6$ $\sigma$). The minimum for N = $50$ occurs around $38$ ps consistent with the trends in the minima for the cases shown in Fig. \ref{7}.(a). The oscillations due to particle-wall collisions seem to occur with different periods consistent with the fact that particles have a distribution of initial positions and speeds.}
\end{figure}
\end{widetext}

features as we dilute the gas keeping the same confined spacing : a slower initial decay, a distinct broadening of the first minimum (Fig.\ref{7}.(a)) as well as a systematic shift of
the minimum to later times (Fig.\ref{7}.(a)) on intermediate time scales, and the emergence of oscillatory behaviour with increasing dilution (N = $50$, Fig.\ref{7}.(b)) on long time scales. At lower dilutions (N = $5000$ and N = $20000$), the minima are not changing with N, where N is total number of particles.\\

In the dilute limit, the particles are expected to bounce between the confining walls ("hard" collisions) resulting in periodic velocity reversals. The VACF would feature oscillations with very little decay as there are very few thermal (random) collisions. As the density increases, the periodicity in the velocity reversals would be disrupted progressively due to increasing thermal (random) collisions on long time scales. Further, as the density increases, wall-mediated collisions would also increase on intermediate and short time scales. Each particle is thus  influenced by a velocity field that is an admixture of thermal collisions and wall-mediated collisions. If the wall-mediated collisions were as random as thermal collisions, the VACF would have decayed rapidly without any other feature. While there is a rapid decay at short timescales, there is a positive correlation that emerges at intermediate timescales.\\

The positive correlations are interpreted as due to the coherent component of the wall-mediated collisions on intermediate timescales. The decay is caused by the incoherent (random) component of the wall-mediated collisions along with purely thermal collisions.
\begin{figure}[H]
\centering
\includegraphics[width=0.5\textwidth]{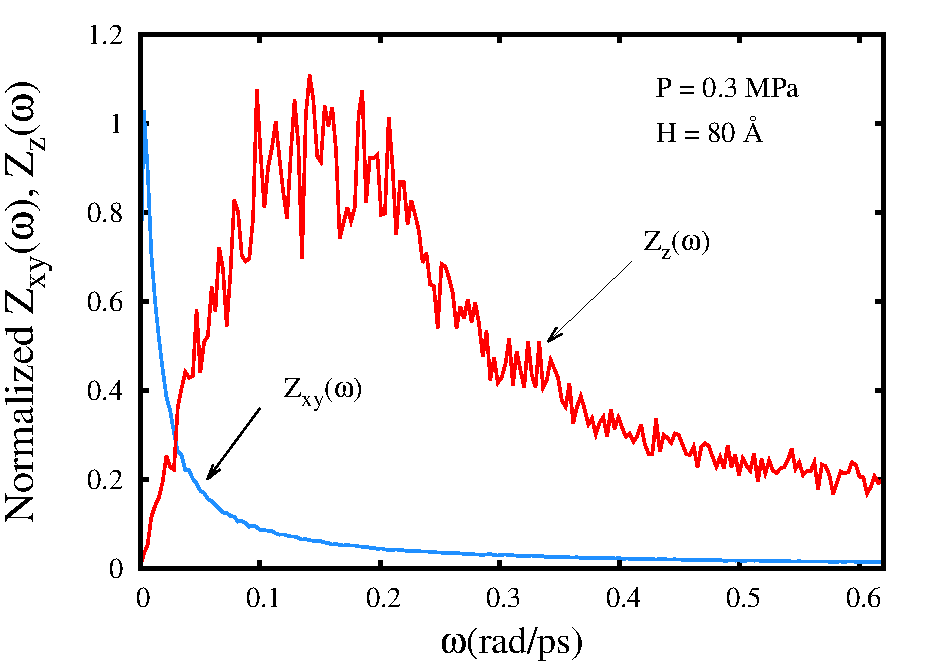}%
\caption{\label{8} Frequency spectra of the normalized VACF$_\perp$ ($Z_{z}(\omega)$) and normalized VACF$_\parallel$ ($Z_{xy}(\omega)$) for H = $80$ $\AA$ ($\approx$ $24$ $\sigma$) at P = $0.3$ MPa.}
\end{figure}

\subsection{\label{sec:level1} Signature of non-diffusive modes}

While the non-diffusive character of the particle dynamics can be noted from the intermediate and long-time scale features of the MSD$_\perp$ and VACF$_\perp$, the Fourier transform of VACF$_\perp$ provides an alternate picture in the frequency domain. The fourier transform ($Z(\omega)$) of normalized VACF ($Z(t)$) is defined as 
\begin{equation}
Z(\omega) = \frac{1}{2 \pi} \int Z(t) exp(-i \omega t) dt
\end{equation}

The Fourier transform of the VACF, denoted by $Z(\omega)$, where $\omega$ is the frequency, is known to yield the density of states (DoS) \cite{Lin}. For a gas phase, $Z(\omega = 0)$ $>$ $0$ and decays monotonically. The non-vanishing of DoS for $\omega$ = 0 corresponds to diffusive modes. For liquids, along with diffusive modes on long time scales $Z(\omega = 0)$ $>$ $0$, there exists non-diffusive modes from caging effects on shorter time scales giving rise to a structure in Z($\omega$) for $\omega$ $>$ $0$ : a local minimum followed by a maximum and decaying thereafter at higher frequencies.\\ 

Fig.\ref{8} shows the Fourier transform of normalized VACF$_\perp$ ($Z_{z}(\omega)$) and normalized VACF$_\parallel$ ($Z_{xy}(\omega)$) for one of the confined systems studied. The DoS for  normalized VACF$_\parallel$ exhibits features of a gas-like system with purely diffusive modes (Z($\omega$ = $0$) $>$ $0$). However, the DoS for normalized VACF$_\perp$ indicates that Z($\omega$ = $0$) = $0$, shows a broad maximum for $\omega$ $>$ $0$ and decays thereafter. The absence of diffusive modes and the presence of the broad maximum, characteristic of non-diffusive modes, indicate that under partial confinement, a gas can exhibit unusual features resembling dense systems in the confining direction. In particular, the broad maximum in the DoS suggests that the non-diffusive modes could be part  of the  vibrational modes, except
\begin{figure}[H]
\centering
\includegraphics[width=0.5\textwidth]{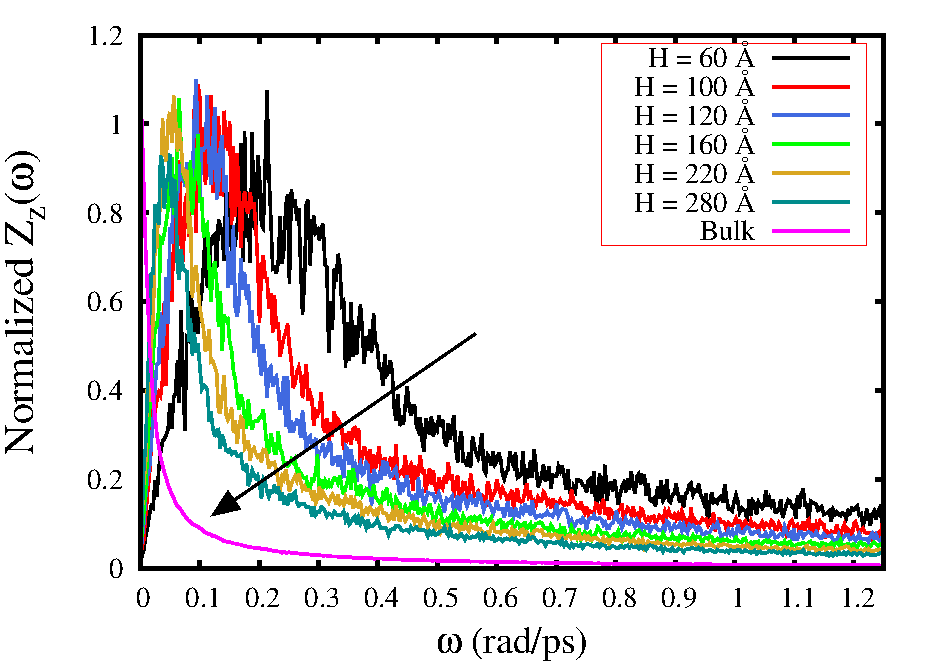}%
\caption{\label{9} Frequency spectra of the normalized VACF$_\perp$ ($Z_{z}$ ($\omega$)) of argon gas for different confined spacings at P = $0.3$ MPa. The arrow indicates the direction of increasing H.}
\end{figure}
 that these vibrational modes seem to be damped in the confined system and not sustained as in a solid phase.\\

To investigate the role of confinement on the frequency spectrum, the Fourier transform of VACF$_\perp$ has been evaluated for several spacings as well as for the bulk and the results are shown in Fig.\ref{9}. The absence of any structure in the DoS for the bulk and the systematic shift of the peak frequency with spacing in the DoS for the partially confined system, establishes that the confining walls contribute to non-diffusive modes. Smaller (larger) confinements lead to the earlier (later) onset of non-diffusive modes, leading to the higher (lower) values for the peak frequencies.

\begin{widetext}
\begin{center}
\begin{figure}[H]
\begin{minipage}{.5\textwidth}
\includegraphics[width=1.0\textwidth]{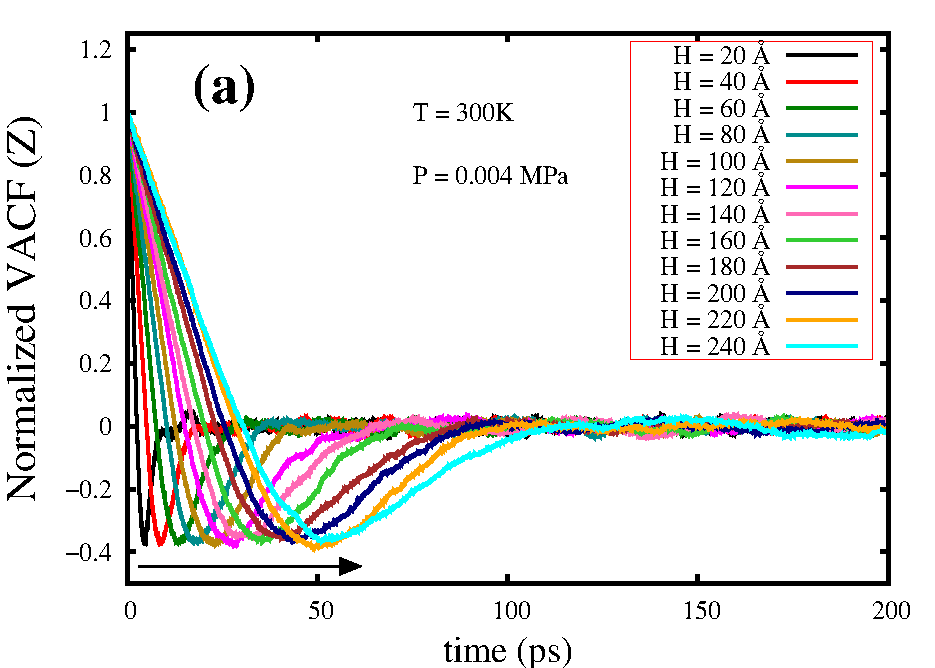}
\end{minipage}
\begin{minipage}{.5\textwidth}
\includegraphics[width=1.0\textwidth]{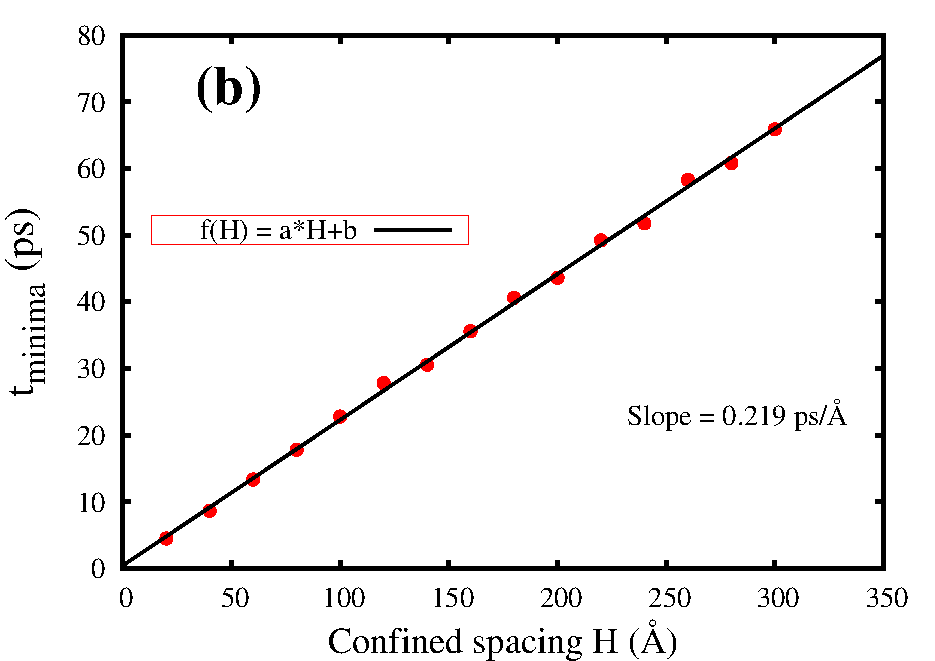}
\end{minipage}
\begin{minipage}{.5\textwidth}
\includegraphics[width=1.0\textwidth]{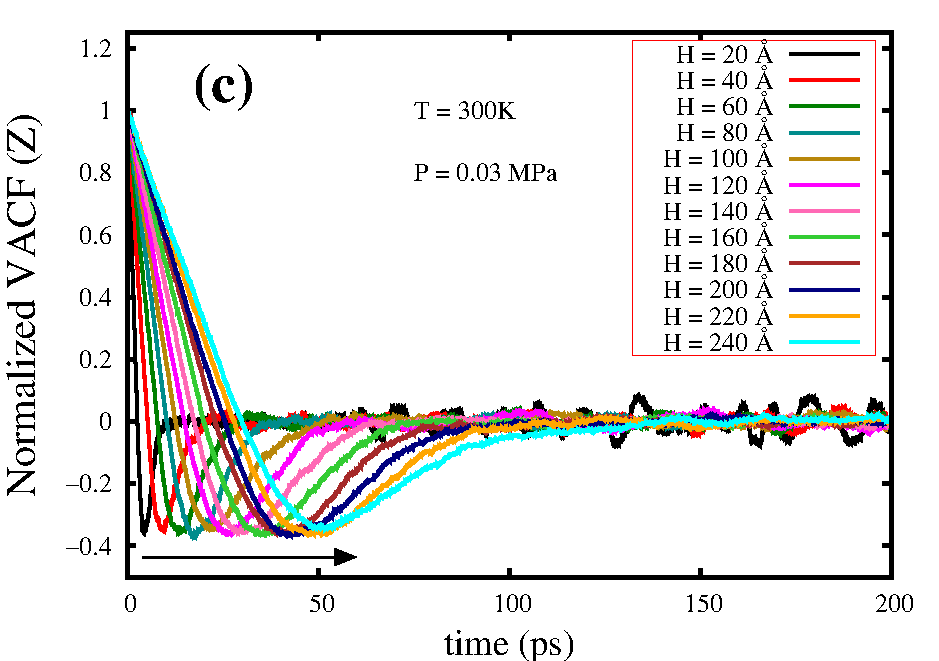}
\end{minipage}
\begin{minipage}{.5\textwidth}
\includegraphics[width=1.0\textwidth]{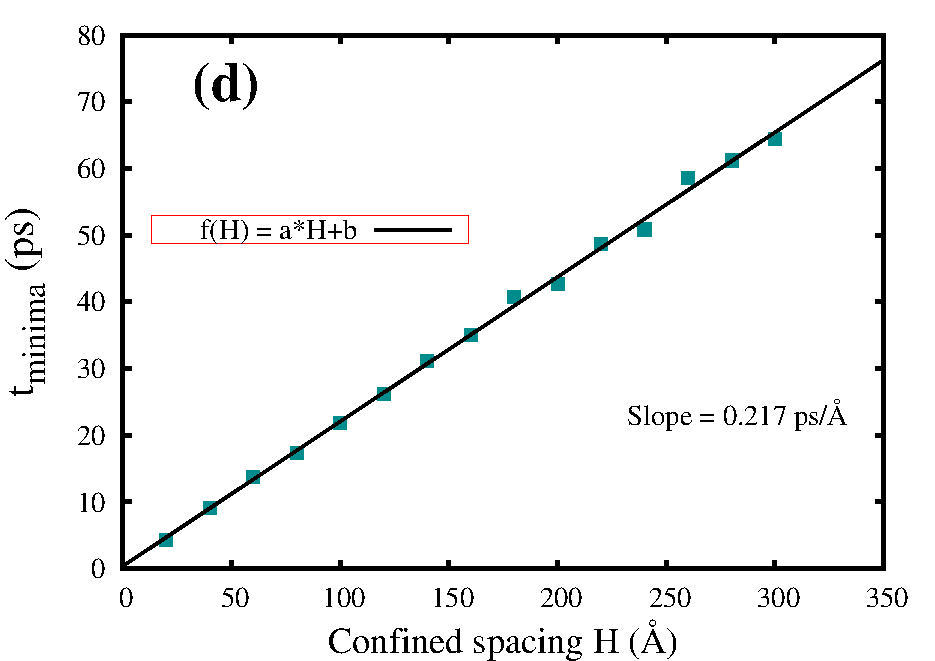}
\end{minipage}
\begin{minipage}{.5\textwidth}
\includegraphics[width=1.0\textwidth]{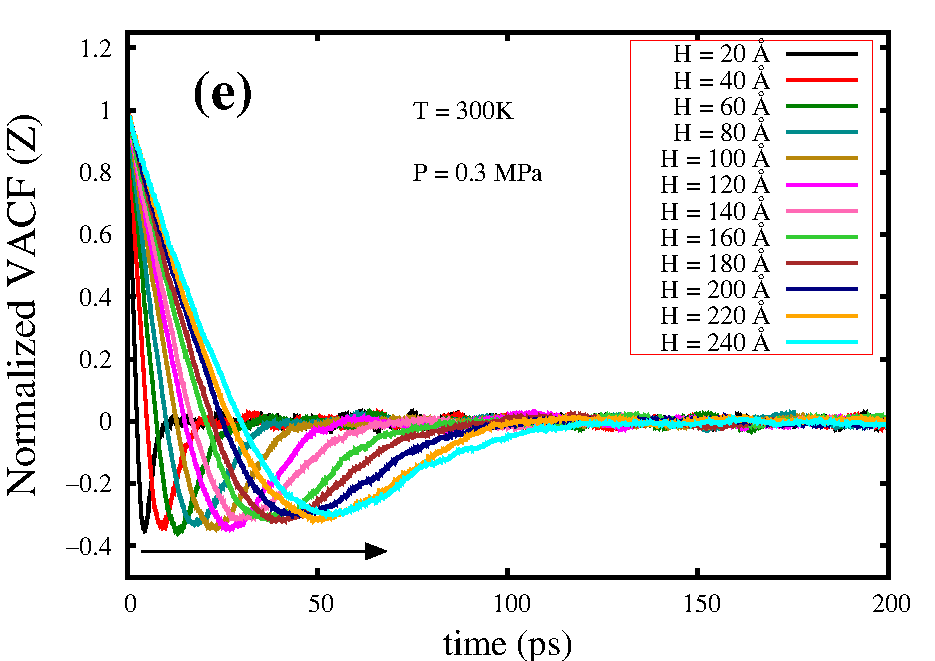}
\end{minipage}
\begin{minipage}{.5\textwidth}
\includegraphics[width=1.0\textwidth]{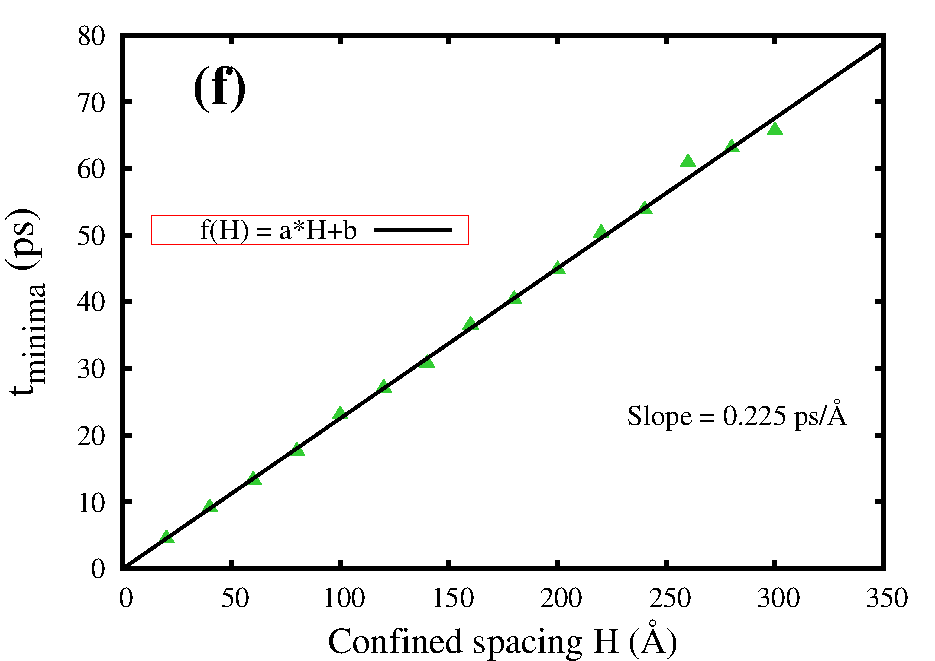}
\end{minipage}
\end{figure}
\end{center}
\end{widetext}
\pagebreak
\begin{widetext}
\begin{center}
\begin{figure}[H]
\begin{minipage}{.5\textwidth}
\includegraphics[width=1.0\textwidth]{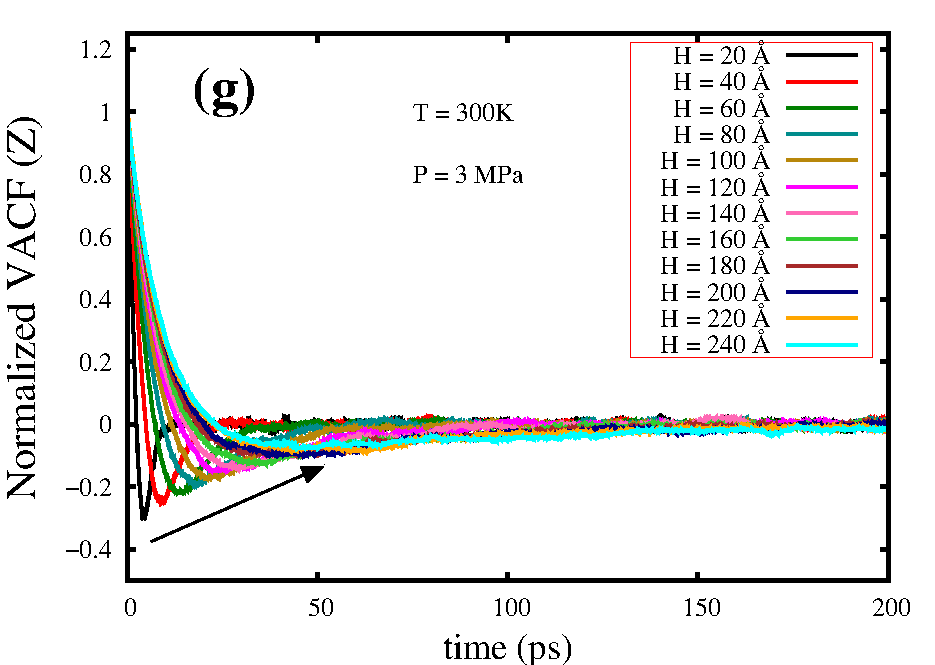}
\end{minipage}
\begin{minipage}{.5\textwidth}
\includegraphics[width=1.0\textwidth]{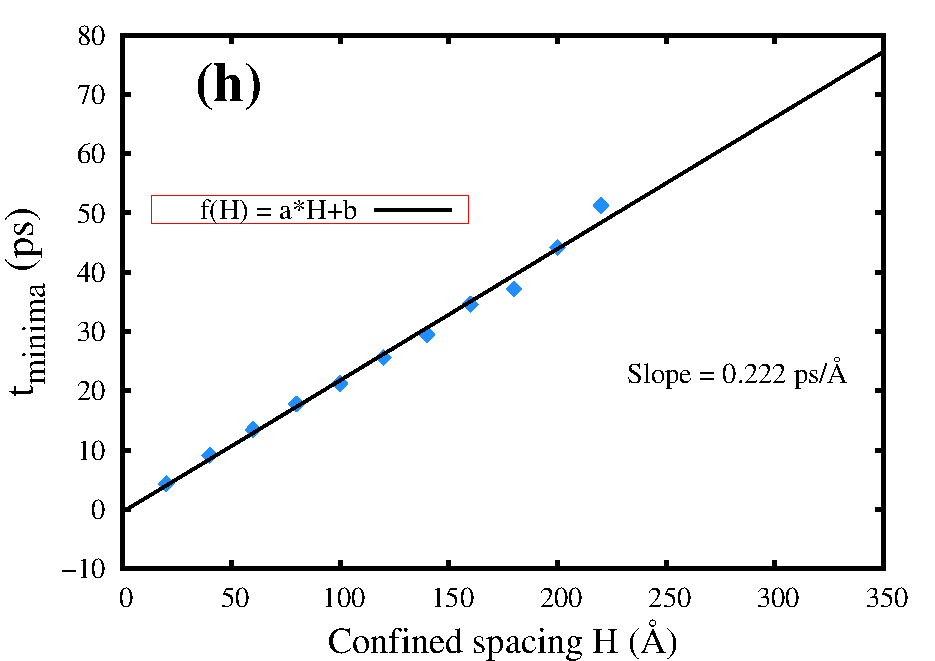}
\end{minipage}
\vspace{0.1cm}
\caption{\label{10} Shifting of $t_{minima}$ at later times with increasing wall-spacing for different P,T state points shown in Fig.\ref{10}.(a), (c), (e), (g). The arrow indicates the direction of increasing H. In Fig.\ref{10}.(b), (d), (f), (h) the linear relations between $t_{minima}$ and the confined separation (H) between the walls have been shown for different P,T state points by linear fit. We observe that the slopes are nearly the same for the entire pressure (/density) regime of study.}
\end{figure}
\end{center}
\vspace{-1cm}
\end{widetext}
\subsection{\label{sec:level1} Connection with speed of sound}

Non-diffusive modes are interpreted as a manifestation of the presence of collective motion.The collective modes conjectured here is related to spontaneous density fluctuations in fluids. In the bulk, these fluctuations are localized in space and time and are distributed statistically throughout the volume. These local density fluctuations propagate very short distances at sound speed, by collective motion, and dissipate subsequently through diffusive motion of individual particles. As there are no restoring forces in dilute fluids, collective motion is not sustained on short lengthscales and timescales. Under partial confinement, however, there is this possibility of the collective motion being reversed by the walls before it dissipates. The reversed collective motion acting on an individual particle, referred to as wall-mediated collisions earlier, is thought to be responsible for the minima in the VACF$_\perp$. We seek further correlations between $H$ and $t_{minima}$ extending the results shown in Fig.\ref{5}.\\

The VACF$_\perp$ has been computed as a function of confined spacing for T = $300$ K over pressures ranging from $3$MPa to as low as $0.004$MPa. This pressure range sets the densities ranging from $48.87$ $kg/m^3$ to $0.064$ $kg/m^3$ at $300$K of argon gas. For P $>$ $5$ MPa argon enters supercritical regime at $300$K. The results are shown in Fig.\ref{10}. In Fig.\ref{10}.(a),(c),(e) it is seen that for every P,T state point in the lower and intermediate pressure regimes, the minima in the VACF$_\perp$ gradually broadens with increasing spacing without much change in their depths. Fig \ref{10}.(g) shows that at higher pressures at $300$K, in addition to broadening, the minima are becoming shallower with increased wall spacing, whereas at lower pressures at the same temperature the depth of the minima are not affected with increased spacing. \\

However, in all cases, a systematic shift of the $t_{minima}$ towards higher values can be seen as spacing is increased. More specifically, a linear relation is observed between the wall spacing and the $t_{minima}$ of the VACF$_\perp$ as shown in Fig.\ref{10}.(b),(d),(f) and (h) for the whole range of pressures(/densities) at $300$K for argon gas. 
\begin{table}[H]
\centering
\caption{\label{table:4} Comparison between average particle velocity along $z$ ($\left\langle v_z \right\rangle_{particle}$) and $H/t_{minima}$ as a function of pressures (P) at $300$K.}
\begin{tabular}{|p{2.5cm}||p{2.7cm}||p{2.7cm}|} 
\hline
& & \\
\textbf{P (MPa)} & \textbf{$H/t_{minima}$ ($\AA/ps$)} & \textbf{$\left\langle v_z \right\rangle_{particle}$ (\textbf{$\AA/ps$})}  \\ [0.5ex]
\hline
& & \\
$0.004$ & $4.57$ & $2.49$   \\
$0.030$ & $4.61$ & $2.48$   \\
$0.3$ & $4.44$ & $2.49$  \\
$3$ & $4.51$ & $2.48$  \\[1ex] 
\hline
\end{tabular}
\end{table}
\onecolumngrid
\begin{widetext}
\begin{center}
\begin{figure}[H]
\begin{minipage}{.5\textwidth}
\includegraphics[width=1.0\textwidth]{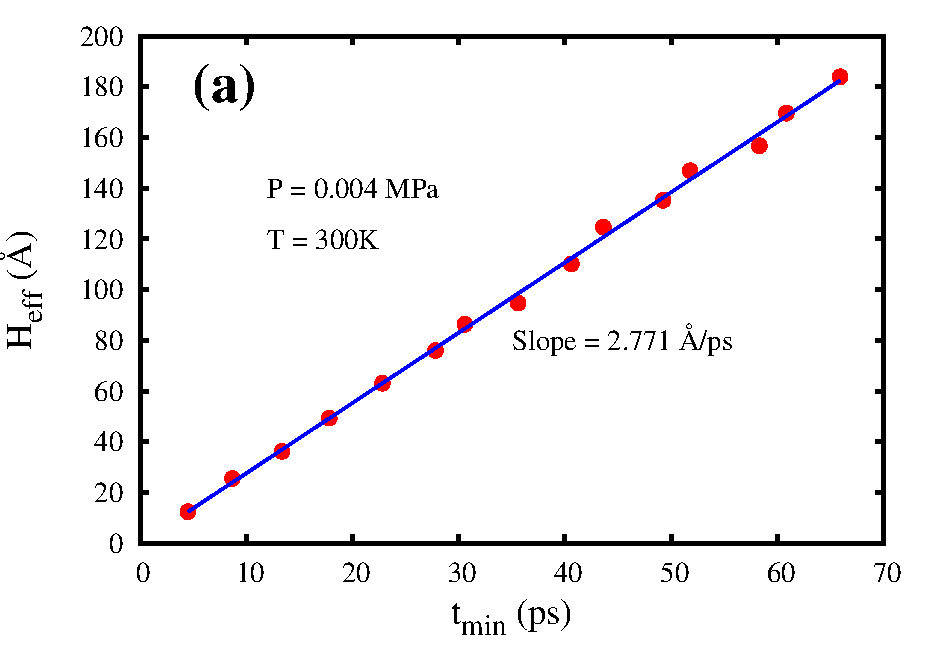}
\end{minipage}
\begin{minipage}{.5\textwidth}
\includegraphics[width=1.0\textwidth]{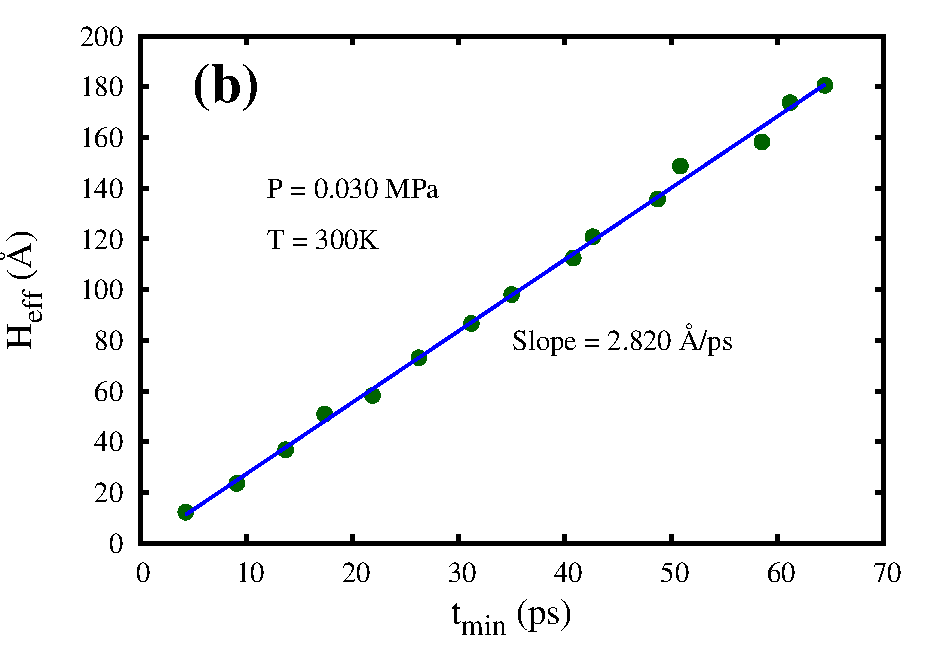}
\end{minipage}
\begin{minipage}{.5\textwidth}
\includegraphics[width=1.0\textwidth]{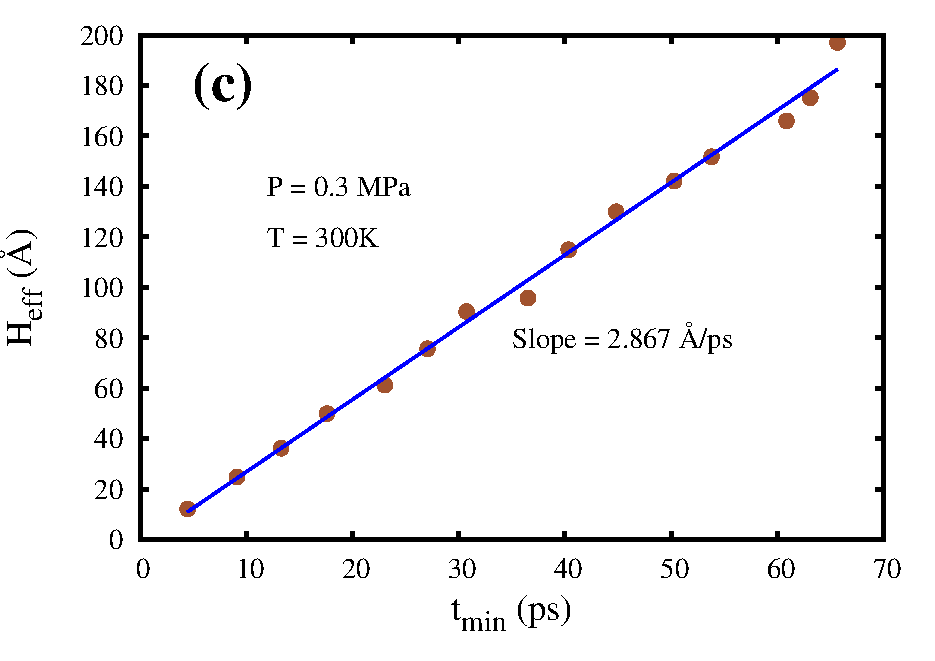}
\end{minipage}
\begin{minipage}{.5\textwidth}
\includegraphics[width=1.0\textwidth]{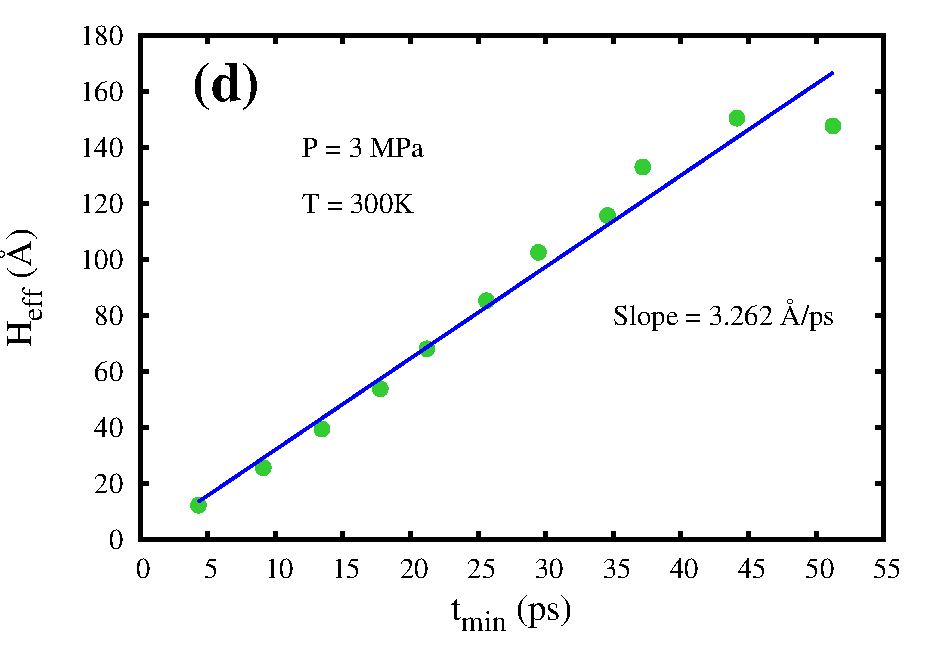}
\end{minipage}
\caption{\label{11} Variation of $H_{eff}$ as a function of $t_{min}$ of VACF$_\perp$ for different state point of argon at $300$K: (a) P = $0.004$ MPa, (b) P = $0.030$ MPa, (c) P = $0.3$ MPa, and (d) P = $3$ MPa. Slopes are extracted from the best linear fit.}
\end{figure}
\end{center}
\vspace{-1cm}
\end{widetext}  

The slopes from these plots for each pressure, having the dimensions of inverse speed, are compared with the particle speed averaged over all the spacings at each pressure in Table \ref{table:4}. The average particle speed is obtained from $\sqrt{a}$, where $a$ is the coefficient of the $t^2$ behaviour of the MSD$_\perp$ in the ballistic regime, for each spacing at a given pressure. The average particle speeds are found to be distinctly lower than reciprocal of the slopes from Fig.\ref{10}.\\

The interpretation that the DoS for VACF$_\perp$ indicates non-diffusive modes, when combined with the observation that the reciprocal of the slopes from Fig. \ref{10}, yield values that are significantly higher than the average particle speeds, suggests that collective motion may be responsible for the minima in VACF$_\perp$. Since particles in a state of ballistic motion cannot be part of the collective motion and since particles in the state of diffusion can only dissipate collective motion, we consider the collective motion to form at $h_1$, defined as the average distance traversed by a particle, with respect to the mid-plane, located somewhere at the beginning of diffusive regime. The sound-wave like excitation thus formed moves at a higher speed towards the wall, undergoes reversal at the wall and traverse back to reach $h_1$ in time $t_{min}$ to "back-scatter" the individual particle. The "back-scatter" appears to critically damp the oscillations of an individual particle as is borne out from the features in MSD$_\perp$ and VACF$_\perp$. The total distance traversed by the sound-wave like excitation in $t_{min}$ is given by $H_{eff}$ = $2\left(\frac{H}{2}-h_1 \right)$. The "back-scatter" is considered to take place at $h_1$. The particle interaction with the reflected excitation marks the end of diffusive regime, which is $H/3$. Therefore we consider the particles not to move appreciable distances ($>$ $h_1$ but $<$ $H/3$) between the generation and subsequent interaction with the excitation. \\

If the sound-wave like excitations were to be responsible for the minima in the VACF$_\perp$ a correlation between $H_{eff}$ and $t_{min}$ would be expected. Indeed, Figure \ref{11} shows a very good linear relation between $H_{eff}$ and $t_{min}$. Since the sound speed is a function of P ( and T: in the present study T is fixed.) the calculations have been performed and presented as a function of P. The role of confinement is examined for each P. The linear relationship found from the study (Fig.11) is consistent with a unique sound speed associated with each pressure. \\ 

The speed of sound estimated from the slope is presented in Table.\ref{table:6}. Noting that the ensemble averaging done to evaluate the VACF$_\perp$ takes into account the fact that particles could have had their origins spatially spread across the width between the walls, the broadening of $VACF_{\perp}-minima$ with increasing wall spacing may be explained as follows: sound waves would, after reflections from the walls, interfere with the particle dynamics at different times (corresponding to the spatial spread seen in MSD$_\perp$) leading to a broadening in the $VACF_{\perp}-minima$. As the wall spacing is reduced, the spatial spread reduces and the temporal spread also reduces leading to a sharper minima in the VACF$_\perp$.\\

The behaviour shown in Fig.\ref{10}.(g), namely minima becoming shallower and broader with increased wall-spacing at P = $3$ MPa, is caused due to the lesser compressibility of the argon gas at this pressure compared to it's lower pressure counterpart. \\

It is known that, at constant pressure and temperature (NPT ensemble) 
the isothermal compressibility is related to the volume fluctuations as
\citep{Allen}
\begin{equation}
\left\langle \delta V^2 \right\rangle_{NPT} = Vk_{B}T\kappa_{T}
\end{equation}
, i.e the lesser compressibility (at higher pressure $3$ MPa) causes lesser volume fluctuations. This, we believe, affects the generation and sustainability of spontaneous sound-wave like excitation at very high pressures under confinement. In other words, the system becomes less compressible to generate significant amplitude of density fluctuations to sustain spontaneous sound-wave like motion at higher pressures. This is consistent with the observations of shallow minima at higher spacings in Fig. \ref{10}.(g) and the scatter in the plot of $H_{eff}$ versus $t_{min}$ at higher spacings in Fig. \ref{11}.(d). \\

We investigate the isothermal compressibility for higher ($3$MPa) and lower pressure ($0.004$MPa) state points as a function of different confined spacings (H). We estimate the number density fluctuations in the form of histograms for LJ fluid (argon) in gaseous regime. We use the relation between isothermal compressibility and density fluctuation \cite{Chandler1968,McQuarrie1976} to find $\kappa_T$.
\begin{equation} \label{eq:8}
\kappa_{T} = \left(\frac{V}{k_{B}T}\right)\frac{\left\langle \left(\Delta \rho_{N}\right)^2\right\rangle}{\rho_{N}^2}
\end{equation}
, where, $\kappa_{T}$ is the isothermal compressibility, $\rho_{N}$ is number density ($\rho_{N}$ = $\frac{N}{V}$), $k_B$ is Boltzmann constant and $\left\langle... \right\rangle$ is the ensemble average. Table.\ref{table:5} contains the values of $\kappa_T$ for different confined spacings. For both higher and lower pressure state points, we observe a lowering of the values of $\kappa_T$ for all the confined spacings, compared to their bulk values. The values of $\kappa_T$ for higher pressure (P = $3$ MPa) under confinement is closer to the bulk $\kappa_T$ value compared to the lower pressure state point (P = $0.004$ MPa). The distinctive difference from the bulk $\kappa_T$ is found to arise from the density fluctuation term $\frac{\left\langle \left(\Delta \rho_{N}\right)^2\right\rangle}{\rho_{N}^2}$ (see equation \ref{eq:8}).\\

While the density remains same for both bulk and confined systems for a particular P,T state point, the density fluctuation is significantly reduced for the confined systems compared to the density fluctuation present in bulk.
\begin{table}[H]
\centering
\caption{\label{table:5} Comparison of isothermal compressibility ($\kappa_T$) and the variation of the square of number density fluctuations $\left(\sigma_{\rho_{N}}^2 \right)$ of bulk and confined argon gas for P = $0.004$ MPa and P = $3$ MPa at $300$ K.}
\begin{tabular}{|p{1.8cm}||p{1.5cm}||p{2.8cm}||p{2.0cm}|} 
\hline
& & & \\
\textbf{P(MPa)} & \textbf{H ($\AA$)} & \textbf{$\sigma_{\rho_{N}}^2$ = $\frac{\left\langle \left(\Delta \rho_{N}\right)^2\right\rangle}{\rho_{N}^2}$} & \textbf{$\kappa_T \left(Pa^{-1}\right)$} \\ [0.5ex]
& & \hspace{1cm}($\times 10^{-5}$) &\\
\hline
& & &\\
$0.004$ & $20$ & $2.80$ & $1.40 \times 10^{-4}$ \\
& $40$ & $2.89$ & $1.44 \times 10^{-4}$ \\
& $60$ & $3.03$ & $1.51 \times 10^{-4}$ \\
& $80$ & $2.91$ & $1.46 \times 10^{-4}$ \\
& $100$ & $2.80$ & $1.40 \times 10^{-4}$ \\
& $120$ & $2.91$ & $1.46 \times 10^{-4}$ \\
& $140$ & $2.88$ & $1.45 \times 10^{-4}$ \\
& $160$ & $2.79$ & $1.39 \times 10^{-4}$ \\
& $180$ & $2.67$ & $1.34 \times 10^{-4}$ \\
& $200$ & $3.03$ & $1.51 \times 10^{-4}$ \\
& $220$ & $2.99$ & $1.50 \times 10^{-4}$ \\
& $240$ & $2.78$ & $1.39 \times 10^{-4}$ \\
& $260$ & $2.89$ & $1.45 \times 10^{-4}$ \\
& $280$ & $2.96$ & $1.48 \times 10^{-4}$ \\
& $300$ & $2.71$ & $1.36 \times 10^{-4}$\\
& Bulk & $4.97$ & $2.49 \times 10^{-4}$\\
\hline
& & &\\
$3$ & $20$ & $3.55$ & $2.48 \times 10^{-7}$ \\
& $40$ & $3.42$ & $2.39 \times 10^{-7}$ \\
& $60$ & $3.6$ & $2.52 \times 10^{-7}$ \\
& $80$ & $4.12$ & $2.89 \times 10^{-7}$ \\
& $100$ & $3.54$ & $2.48 \times 10^{-7}$ \\
& $120$ & $3.6$ & $2.52 \times 10^{-7}$ \\
& $140$ & $3.47$ & $2.43 \times 10^{-7}$ \\
& $160$ & $3.76$ & $2.63 \times 10^{-7}$ \\
& $180$ & $3.54$ & $2.48 \times 10^{-7}$ \\
& $200$ & $3.68$ & $2.58 \times 10^{-7}$\\
& $220$ & $3.6$ & $2.52 \times 10^{-7}$\\
& $240$ & $3.65$ & $2.56 \times 10^{-7}$\\
& $260$ & $3.25$ & $2.28 \times 10^{-7}$\\
& $280$ & $3.33$ & $2.33 \times 10^{-7}$\\
& $300$ & $3.07$ & $2.15 \times 10^{-7}$\\
& Bulk & $4.61$ & $3.23 \times 10^{-7}$\\ [1ex] 
\hline
\end{tabular}
\end{table}
Further, the trends in Table.\ref{table:5} indicate that the density fluctuations are consistently smaller (larger) for the higher (lower) pressure over the range of confinement spacings studied, as expected. The system at the maximum spacing is still far from the bulk as can be seen from Table \ref{table:5}.\\

We also note from Table.\ref{table:5} that the number density fluctuations in bulk are not significantly different for higher and lower pressure cases. The dominant contributions for $\kappa_T$ for bulk gas come from the $\left(\frac{V}{k_{B}T} \right)$ term, which is very different for different pressures at fixed temperature ($300$ K).\\

For the highest pressure under study ($3$ MPa), even under confinement, sound speed is found to be close to the NIST \cite{NIST} bulk phase value (Table.\ref{table:6}). This suggests the dominant nature of particle-particle collisions at higher pressures. As we go from higher to lower pressure P,T points gradually, the effect of confinement becomes stronger and the value of sound speed starts to decrease with respect to NIST bulk sound speed at the corresponding pressure at $300$ K (Table.\ref{table:6}).\\

To ensure that estimates made from MD simulations are reasonable and reliable, we carry out MD simulations to calculate $c_s$ for the bulk phase of argon gas at each of these four P,T state points from the isothermal compressibility ($\kappa_T$) defined in Eq.\ref{eq:8}. We use the thermodynamic relation between adiabatic and isothermal compressibility for mono-atomic gas as $\kappa_{T}$ = $\gamma \kappa_{S}$ and calculate the speed of sound from the well known relation \cite{Robert1976} $c_s$ = $\frac{1}{\sqrt{\rho \kappa_{S}}}$, where $c_s$ is the speed of sound, $\rho$ is the mass density, $\gamma$ is heat capacity ratio and $\kappa_S$ is the adiabatic compressibility. The values, tabulated in Table.\ref{table:6} can be seen to agree fairly well with that of the NIST bulk phase data \cite{NIST} for sound speed.\\
 
The sound speed estimates using the slopes in Fig.\ref{11}, based on MSD$_\perp$, are presented in Table.\ref{table:6} over a wide range of densities (about $4$ orders of magnitude). 
\begin{table}[H]
\centering
\caption{\label{table:6} Comparison of the variation of speed of sound($c_s$) of argon in the gas regime at $300$K for different densities($\rho$) calculated and estimated from MD simulation and that of the NIST values.}
\begin{tabular}{|p{1.3cm}||p{1.5cm}||p{1.7cm}||p{1.7cm}||p{1.0cm}|} 
\hline
& & \multicolumn{3}{|c|}{\textbf{Sound speed($c_{s}$) ($m/s$)}} \\
& & & & \\
\hline
\textbf{P(MPa)} & \textbf{$\rho$($kg/m^3$)} & \textbf{Estimated from slope (MD)} &  \textbf{Computed via bulk $\kappa_T$ (MD)} & \textbf{NIST} \\ [0.5ex]
\hline
& & & &\\
$0.004$ & $0.064$ & $277.1$ & $324.95$ & $322.6$ \\
$0.03$ & $0.48$ & $282.04$ & $319.79$  & $322.62$ \\
$0.3$ & $4.8$ & $286.71$ & $325.54$ & $322.83$ \\
$3$ & $48.87$ & $326.2$ & $337.47$ & $325.58$ \\[1ex] 
\hline
\end{tabular}
\end{table}
These estimates can be seen not to agree entirely with the NIST \cite{NIST} bulk phase data. Further, sound speed estimates from standard expressions $c_s$ = $\frac{1}{\sqrt{\rho \kappa_{S}}}$ and $\kappa_T$ = $\gamma \kappa_s$, and using $\kappa_T$ from the fourth column of Table.\ref{table:5}, leads to consistently higher values than the NIST bulk phase data.    These discrepencies may be indicative of the need to re-examine the standard relation $\kappa_T$ = $\gamma \kappa_s$ and perhaps $c_s$ = $\frac{1}{\sqrt{\rho \kappa_{S}}}$ under strong confinement.\\

\section{\label{sec:4}Summary and Conclusions}

We have carried out MD simulations of a confined gas to investigate the role of partial confinement, effected by two plane parallel reflective walls, on the particle dynamics. The study, done using LAMMPS, considered a system of $20000$ particles both in bulk and in partial confinement, interacting via the Lennard-Jones potential, at T = $300$K and pressures ranging from $0.004$ MPa to $3$ MPa simulating a wide range of densities. The confinement spacing ranged from $20$ $\AA$ ($\approx$ 6 $\sigma$) to $300$ $\AA$ ($\approx$ $90$ $\sigma$).\\

Confinement changes MSD$_\perp$ and VACF$_\perp$ strongly. It has very little effect on MSD$_\parallel$ and VACF$_\parallel$. At short time scales (ballistic regime) MSD$_\perp$ is barely influenced by confinement. Over intermediate time scales (diffusive regime), the diffusion coefficient ($D_{\perp}$) is significantly reduced, with respect to the bulk, due to confinement. Over longer time scales, confinement effects are the strongest leading to sub-diffusive motion. The transition from diffusive to sub-diffusive behaviour correlates with the wall spacing ($\approx$ $\frac{1}{3}$ of wall spacing for P = $0.004$ MPa, $0.03$ MPa, $0.3$ MPa and $\approx$ $\frac{1}{3.9}$ for P = $3$ MPa). Irrespective of pressure, we observe for all P,T state points and for all the spacings studied, the asymptotic values for $\sqrt{MSD_{\perp}}$ turns out to be $\frac{H}{2.45}$.\\ 

A well-defined time-window is found that marks the end of ballistic regime and the beginning of the diffusive regime in the bulk as well as under confinement. For small wall spacing, the time-window is small. It is worth noting that experiments in the bulk confirms the existence of this time-window \cite{Huang,Tongcang} albeit for Brownian particles. \\

It may be noted that unlike the particle dynamics in the gas phase studied in this article, particle dynamics  parallel to the walls, have been found to be affected in the dense liquid phase \cite{Vadhana}, and in the glass forming phase \cite{Suvendu}.\\
 
VACF$_\perp$ shows a rapid decay corresponding to the diffusive regime and exhibits a well-defined minimum over time scales corresponding to the transition from diffusive to sub-diffusive regimes and nearly vanishes in the sub-diffusive regime. The Fourier transform of VACF, related directly to the density of states (DoS), indicates the presence of non-diffusive modes in the spectrum of normalized  VACF$_\perp$ and diffusive modes in the spectrum of normalized VACF$_\parallel$ . The frequencies at which maxima in the DoS occur are found to move closer to zero as the wall spacing is increased confirming the role of confinement in sustaining the non-diffusive modes.\\

Reducing the contributions of the thermal collisions, by diluting the gas further, reveals the presence of sustained velocity reversals induced by wall-mediated collisions. The results confirm the significant role played by non-thermal collisions in controlling the dynamics of the particles under confinement over intermediate and long time scales.\\

It is established from the strong linear relationship between the wall spacing and the time at which the VACF$_\perp$-minima occurs, that there exists a phenomenon involving a higher speed than the average particle speed. It is conjectured that collective motion triggered at the early stages of the diffusive regime propagates at the higher speed, gets reversed at the wall and acts back on the particle leading to the minima in the VACF$_\perp$. Using the characteristic distance associated with the starting of the diffusive motion, and the $t_{min}$ associated with VACF$_\perp$, the speed of this collective excitation regarded as sound speed, is estimated for several pressures spread over the whole range considered in this study. \\  

These features, particularly the connection with speed of sound, combined with the observation that the compressibility has been found to be higher in the low pressure regimes, when compared with that in the high pressure regimes, suggest that  correlated motion resembling sound waves can be spontaneously generated and sustained at low pressure regimes better than at the higher pressure regimes.\\ 
  
Although the possibility of sound-wave generation and reflection from confining boundaries has been reported in the literature for liquids \cite{Magda1985,Pagonabarraga1999}, the present study has systematically investigated the effects of partial confinements for a gas-like phase, established strong correlation between MSD$_\perp$ and VACF$_\perp$ and shown the possibility of spontaneous generation and propagation of sound-wave like disturbances normal to the walls.\\

The density fluctuations seem to be considerably influenced by the confinement even though they are over larger length scales than the length scales associated with particle-particle and particle-wall interactions. Further, the sound speed estimates made in the present study appear not to agree entirely with the NIST bulk phase data. It would be interesting to conduct experiments to validate these findings. Further there appears a need to re-examine the relation $\kappa_T$ = $\gamma$ $\kappa_s$ and perhaps $c_s$ = $\frac{1}{\sqrt{\rho \kappa_s}}$ for strongly confined systems.\\





\begin{acknowledgments}
We sincerely thank the Reviewers for their critical assessment of the manuscript. We acknowledge the help of HPCE,IIT Madras for high performance computing. KG expresses his gratitude to Department of Science and Technology(DST), the Government of India for providing INSPIRE Fellowship.  
\end{acknowledgments}

\bibliographystyle{apsrev4-1}
\nocite{*}

\providecommand{\noopsort}[1]{}\providecommand{\singleletter}[1]{#1}%
\end{document}